\documentclass[%
 aip,
 amsmath,amssymb,
 reprint,%
]{revtex4-1}
\usepackage[utf8]{inputenc}
\usepackage{graphicx}
\usepackage{dcolumn}
\usepackage{bm}
\usepackage{breqn}
\usepackage[utf8]{inputenc}
\usepackage[T1]{fontenc}
\usepackage{mathptmx}
\usepackage{etoolbox}
\usepackage{booktabs}
\usepackage{color}
\usepackage{xcolor}
\makeatletter
\def\@email#1#2{%
 \endgroup
 \patchcmd{\titleblock@produce}
  {\frontmatter@RRAPformat}
  {\frontmatter@RRAPformat{\produce@RRAP{*#1\href{mailto:#2}{#2}}}\frontmatter@RRAPformat}
  {}{}
}%
\makeatother
\begin{document}

\preprint{AIP/123-QED}

\title{N-Component Free Energy Lattice Boltzmann Method with Reduction Consistency and Global Momentum Conservation}
\author{Michael Rennick}%
\affiliation{$Institute\ for\ Multiscale\ Thermofluids,\ School\ of Engineering,\ University\ of\ Edinburgh,\ Edinburgh\ EH9\ 3FD,\ United\ Kingdom$}
\author{Xitong Zhang}%
\affiliation{$Institute\ for\ Multiscale\ Thermofluids,\ School\ of Engineering,\ University\ of\ Edinburgh,\ Edinburgh\ EH9\ 3FD,\ United\ Kingdom$}
\author{Tim Niklas Bingert}%
\affiliation{$Institute\ for\ Mechanical\ Process\ Engineering\ and\ Mechanics,\ Karlsruhe\ Institute\ of\ Technology,\ Karlsruhe,\ Baden-W{\ddot{u}}rttemberg, Germany$}
\affiliation{$Lattice\ Boltzmann\ Research\ Group,\ Karlsruhe\ Institute\ of\ Technology,\ Karlsruhe,\ Baden-W{\ddot{u}}rttemberg, Germany$}
\author{Mathias J. Krause}
\affiliation{$Institute\ for\ Mechanical\ Process\ Engineering\ and\ Mechanics,\ Karlsruhe\ Institute\ of\ Technology,\ Karlsruhe,\ Baden-W{\ddot{u}}rttemberg, Germany$}
\affiliation{$Lattice\ Boltzmann\ Research\ Group,\ Karlsruhe\ Institute\ of\ Technology,\ Karlsruhe,\ Baden-W{\ddot{u}}rttemberg, Germany$}
\author{Halim Kusumaatmaja}
\email{halim.kusumaatmaja@ed.ac.uk}
\affiliation{$Institute\ for\ Multiscale\ Thermofluids,\ School\ of Engineering,\ University\ of\ Edinburgh,\ Edinburgh\ EH9\ 3FD,\ United\ Kingdom$}

\date{\today}

\begin{abstract}
We present a free energy lattice Boltzmann model capable of simulating fluid systems with an arbitrary number of immiscible components in principle. Our method is strictly reduction consistent, ensuring that absent fluid components do not spontaneously nucleate. We introduce a novel discretization of the surface tension force that globally conserves momentum to machine precision, and we enforce reduction consistency through a flux correction that is independent of the mobility.
The method is benchmarked with a range of static and dynamic problems, including: liquid lenses, Janus droplets, quaternary phase separation, and six-component layered Poiseuille flow, and we obtain excellent agreement with theoretical predictions throughout. Finally, we demonstrate the applicability of the proposed method through patterned liquid surfaces and microfluidic emulsion droplet generation.
\end{abstract}

\maketitle

\section{Introduction}
Multicomponent fluid flows underpin a broad range of natural and industrial applications, and there is a recent drive to study systems composed of three or more immiscible components. For example, emulsion droplets consisting of multiple nested phases offer promising platforms for drug delivery and for controlled chemical reactions between their constituent fluids~\cite{microshear,microemulsion1,Janusmicrofluidic,microemulsion2,microemulsion3,microemulsion4,microfluidic5}. Patterned liquid infused surfaces employ multiple lubricating fluids to direct droplet motion, while maintaining an ultra smooth lubricant surface and preventing pinning on surface heterogeneities~\cite{xitong,pal1,pal2}. In biological contexts, phase separation of proteins and nucleic acids within cells can lead to a large number of distinct condensates~\cite{zwicker,phasesepmany1,phasesepmany2,phasesepcondensate}, and engineered DNA nanostar systems~\cite{Abraham2024-io,Fabrini2024} have been used to generate up to nine immiscible phases~\cite{chaderjian2025diversedistinctdenselypacked}.

A wide range of numerical techniques have been developed to capture systems comprising two, and in some cases three, immiscible fluids, including sharp~\cite{BRACKBILL1992335,SUSSMAN2007469,ABUALSAUD2018896} and diffuse~\cite{diffuse1,threecomp,diffuse2} interface approaches. However, methods capable of handling an arbitrary number ($N$) of distinct phases in principle are still limited. The principal challenge lies in constructing a system of macroscopic equations that is reduction consistent, where an $N$ component system exactly reproduces the corresponding $N-1$ component system when one fluid is absent~\cite{boyerncomp,dongncomp}. Without this property, absent fluid components can spontaneously nucleate, severely compromising simulation accuracy and stability.

In this work, we formulate a continuum model for $N$-component fluids that is full reduction consistent, and develop a suitable free energy lattice Boltzmann method (LBM)~\cite{lbmbook1,lbmbook2} to solve it.
The LBM is widely used to study multicomponent fluid problems~\cite{lbmapplication1,lbmapplication2,lbmapplication3}, and is becoming increasingly popular due to its exceptional computational efficiency in large, complex geometries~\cite{Raeli2025,kummerlander_2025_17899765,olbPaper2021}. Moreover, as a diffuse interface method, interface motion, including break-up and coalescence events, can be tracked implicitly from the evolution of a continuous phase field. A number of these  approaches have been developed to study multicomponent flow problems with up to three fluids~\cite{diffuse2,lbmapplication3,ALamura_1999, abadi_ternary_2018}, but lattice Boltzmann methods capable of resolving fluid flows with four or more distinct fluid phases are limited. In addition, some approaches have been developed to resolve large numbers of immiscible droplets of the same fluid~\cite{TIRIBOCCHI20251,Montessori_Lauricella_Tirelli_Succi_2019}, preventing coalescence using near contact forces. While these techniques have the advantage of remaining computationally feasible for hundreds or thousands droplets, they do not allow for independent selection of the interfacial tensions between fluid phases, and rely on interfaces that are phase separated a priori. Our method is capable of simulating fluid flows of multiple immiscible fluid components with independent selection of all interfacial tensions between the fluids.

In contrast to previous LBM approaches, our method has two key advantages. First, in order to capture both the interfacial dynamics and the diffusive phase separation physics, we evolve the fluid concentrations using a system of Cahn-Hilliard equations. Previous attempts to incorporate reduction consistency in Cahn-Hilliard systems have typically relied upon a specific degenerate form of the diffusive mobility~\cite{dongncomp,lbmncomp1,lbmncomp2}, which leads to zero flux for an absent fluid. However, as a consequence, the diffusive physics is restricted by the approach for reduction consistency. This is a serious limitation, as we will exemplify here in the context of liquid-liquid phase separation studies. Instead, our strategy is to apply a source term in the Cahn-Hilliard equations to cancel any erroneous flux, while the mobility is a free parameter to tune the diffusivity~\cite{boyerncomp}. We demonstrate that the model reproduces established hydrodynamic and diffusive scaling laws for phase separation~\cite{phasesepscaling,phasesepscalingsource,Wagner_2001,WagnerYeomans}. Second, previous methods utilise a discretisation of the potential form of the surface tension force~\cite{lbmbook1,lbmncomp1,lbmncomp2}, which is not strictly momentum conserving and can induce a significant whole domain drift. Here, we resolve this issue by deriving a momentum conserving discretisation of the surface tension force that eliminates this drift to machine precision.

This paper is structured as follows. In Sec. II, we present the continuum model solved in this work, including the approach to ensure reduction consistency. Then, in Sec. III, we provide the LBM scheme and conservative discretisation of the surface tension force. We then provide a series of systematic static and dynamic benchmark tests in Sec. IV to validate our model, including (i) liquid lenses and Janus droplets, (ii) hydrodynamic and diffusive phase separation, and (iii) Poiseuille flow with six fluid layers.
To illustrate the capabilities of the method, in Sec. V we showcase applications to study patterned liquid surfaces and a microfluidic design for generating emulsion droplets.
Finally, we conclude our work in Sec. VI.

\section{Continuum Model}\label{sec:continuum}

We derive our method from three distinct continuum equations. First and second, we solve the incompressible continuity and Navier-Stokes equations to capture the incompressibility constraint and momentum conservation respectively
\begin{eqnarray}
& \mathbf{\nabla}\cdot\mathbf{v} = 0,
\label{eqn:ce} \\
& \partial_t(\rho \mathbf{v}) + \mathbf{\nabla}\cdot(\rho\mathbf{vv})= - \mathbf{\nabla}P + \eta\Delta \mathbf{v}+\mathbf{F}.
\label{eqn:nse}
\end{eqnarray}
In these equations, $\mathbf{v}$, $\rho$, $\eta$ and $P$ are the fluid velocity, density, dynamic viscosity and hydrodynamic pressure. $\mathbf{F}=\mathbf{F}_s+\mathbf{F}_b$ includes the surface tension force $\mathbf{F}_s$ and any external body force $\mathbf{F}_b$.

Thirdly, to track the evolution of the interface locations for $N$ immiscible components, we evolve $N-1$ Cahn-Hilliard equations for the fluid volume fractions $C_i$
\begin{equation}
\label{eq:CHFlux}
\partial_t C_i+\nabla\cdot(\mathbf{v}C_i)=-\nabla\cdot\mathbf{J}_i,\ \ \forall i \in \{1,2,\dots,N-1\},
\end{equation}
with the evolution of $C_N$ implicit from the constraint that $\sum^N_{i=1} C_i=1$. The fluid density $\rho$ can then be calculated from the volume fractions and component densities $\rho_i$ as
\begin{equation}
\rho=\sum_{i=1}^N\rho_iC_i,
\end{equation}
although we fix $\rho_i=1$ in this study for simplicity, and consider cases where inertial effects are negligible.

The surface tension force $\mathbf{F}_s$ and diffusive flux $\mathbf{J}_i$ can be derived in terms of the fluid free energy $\mathcal{F}$. The following form allows independent selection of the surface tensions between each of $N$ components~\cite{boyerncomp,dongncomp}
\begin{align}
\label{eqn:FreeEnergy}
\mathcal{F} &= \int \left[ E_B + E_I \right] dV, \\
E_B &= \sum_{i\neq j}^N \frac{\beta_{ij}}{2} \Big[ f(C_i) + f(C_j) - f(C_i + C_j) \Big], \notag \\
f(C_i) &= C_i^2 (1-C_i)^2, \quad E_I = -\sum_{i\neq j}^N \frac{\kappa_{ij}}{2} \nabla C_i \cdot \nabla C_j. \notag
\end{align}
The bulk term $E_B$ ensures that, far from the interface, the energy is minimised when $C_i$ is either $1$ or $0$ for each component. The gradient term $E_I$ leads to diffuse interfaces with characteristic width $D=\sqrt{8\kappa_{ij}/\beta_{ij}}$ and ensures that the excess free energy per unit area of the diffuse interface corresponds to the surface tensions $\sigma_{ij}=\sqrt{2\kappa_{ij}\beta_{ij}/9}$. In this study, we set $D=4$.

A common form of the diffusive flux is~\cite{boyerncomp,dongncomp,lbmncomp1,lbmncomp2}
\begin{equation}
\mathbf{J}_i = M\mathbf{\nabla} \left(\sum^N_{j=1}\alpha_{ij}\mu_j\right),
\label{eq:standardflux}
\end{equation}
where $M$ controls the diffusive mobility and $\mu_j=\delta\mathcal{F}/\delta C_j$ is the chemical potential of fluid $j$
\begin{align}
\mu_j = \sum_{k,k\neq j}^N\Biggl[& \beta_{jk} \Big( f'(C_j) - f'(C_j + C_k) \Big)\\
&+\kappa_{jk} \nabla^2 C_k\Biggr],\notag  \\
f'(C_j) = 2C_j(&1 - 3C_j + 2 C_j^2)\notag.
\label{eq:mu}
\end{align}
The coefficients $\alpha_{ij}$ are given by solving the linear system~\cite{boyerncomp}
\begin{equation}
\boldsymbol{\alpha\sigma}=\boldsymbol{I}+\boldsymbol{\gamma}\otimes\boldsymbol{1}^T,
\label{eq:alpha1}
\end{equation}
\begin{equation}
\boldsymbol{\alpha}\boldsymbol{1}=\boldsymbol{0},
\label{eq:alpha2}
\end{equation}
where $\boldsymbol{\gamma}$ is a vector of size $N$ and $\boldsymbol{\sigma}$ is the symmetric matrix of surface tensions. Additionally, $\otimes$ represents the Kronecker product and $\boldsymbol{1}$ and $\boldsymbol{0}$ are vectors of size $N$ with each entry as $1$ and $0$ respectively. To ensure reduction consistency, the diffusion flux $\mathbf{J}_i$ of an absent fluid component $i$ must be zero. In general, this is not met for arbitrary choices of $M$ and $\alpha_{ij}$, except for $N=2$.

\begin{figure}[t!]
    \centering
	\includegraphics[width=\columnwidth]{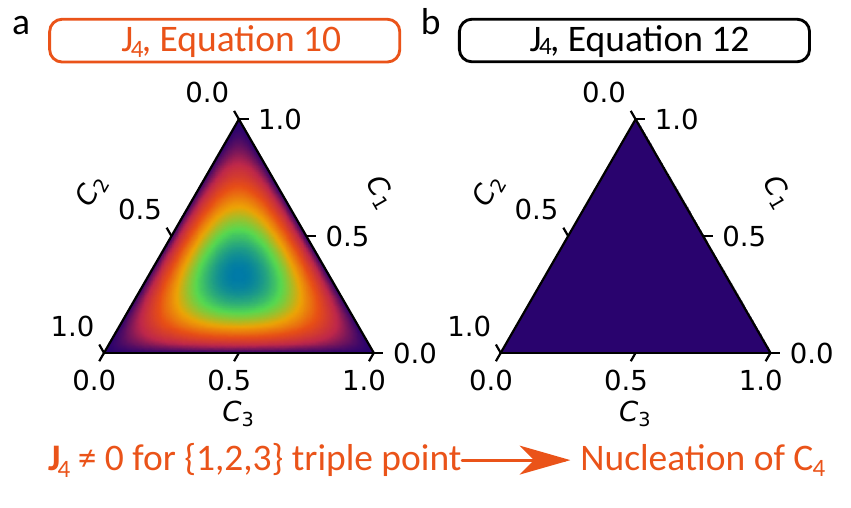}
	\caption{(a) Bulk flux from Eq.~\ref{eq:fluxwrong} in the Cahn-Hilliard equation for $C_4=0$ as a function of $C_1$, $C_2$ and $C_3$. The non-zero flux in this case leads to the nucleation of $C_4$. (b) Bulk flux from Eq.~\ref{eq:MU} in the Cahn-Hilliard equation for $C_4=0$ as a function of $C_1$, $C_2$ and $C_3$, demonstrating that it provides reduction consistency.}
	\label{fig:ExcessFlux}
\end{figure}

To identify the root cause of reduction inconsistency, we can calculate the resulting bulk flux in the Cahn-Hilliard equation for the $N>3$ component system as
$\mathbf{J}_{i,\mathrm{bulk}}=-M\nabla\sum^N_{j=1}\alpha_{ij}\mu_{j,b}$, where $\mu_{j,b}=\partial E_B/\partial C_j$ and $\alpha_{ij}$ satisfies Eq.~\ref{eq:alpha1} and Eq.~\ref{eq:alpha2}. We ignore the gradient terms in $\mu_j$ as these are naturally reduction consistent. Using this approach, we obtain~\cite{boyerncomp}
\begin{align}
\mathbf{J}_{i,\mathrm{bulk}} 
&= \frac{12M}{D}\nabla\Biggl[
   -C_i(1 - 3C_i + 2 C_i^2) \notag \\ 
&+ 6\sum_{1\leq j< k< l\leq N}
   \Bigl(\alpha_{ij}(\sigma_{jk}+\sigma_{jl})
       +\alpha_{ik}(\sigma_{jk}+\sigma_{kl})\notag
       \\ 
&+\alpha_{il}(\sigma_{jl}+\sigma_{kl})
       -\gamma_i\Bigr) C_jC_kC_l
   \Biggr].
\label{eq:fluxwrong}
\end{align}
Fig.~\ref{fig:ExcessFlux} (a) illustrates the case where $N = 4$ and $C_4 = 0$ initially. However, employing Eq.~\ref{eq:fluxwrong}, we find that $\mathbf{J}_{4,\mathrm{bulk}} \neq 0$ if at least three other components are present, as $C_jC_kC_l\neq0$. This leads to unphysical nucleation of $C_4$ at any triple point formed of other fluids.

To resolve this issue, we note that, for three component systems, the flux corresponds to
\begin{align}
\mathbf{J}_{i,\mathrm{bulk}} 
&= \frac{12M}{D}\nabla\Biggl[
   -C_i(1 - 3C_i + 2 C_i^2) \notag \\ 
&+6\Bigl(\alpha_{ii}(\sigma_{ij}+\sigma_{ik})
       +\alpha_{ij}(\sigma_{ij}+\sigma_{jk})\notag
       \\ 
&+\alpha_{ik}(\sigma_{ik}+\sigma_{jk})
       -\gamma_i\Bigr) C_iC_jC_k
   \Biggr],\quad 1\leq i<j<k\leq 3.
\label{eq:fluxright}
\end{align}
For $N=3$, reduction consistency is guaranteed by the constraint that $\alpha_{ij}$ satisfies Eq.~\ref{eq:alpha1} and Eq.~\ref{eq:alpha2}, and the flux in Eq.~\ref{eq:fluxright} is zero when $C_i=0$. However, for $N > 3$, this is not possible using the expression in Eq.~\ref{eq:fluxwrong}, as we do not have sufficient independent choices for the $\alpha_{ij}$ terms to also cancel the $C_jC_kC_l$ terms, which are in general non-zero when $C_i=0$.

Correspondingly, our strategy is to recast the flux such that the $N$ component flux must reduce to Eq.~\ref{eq:fluxright} when only three fluids are present. This is given by
\begin{equation} 
\mathbf{J}_i = M\mathbf{\nabla} \left(\sum^N_{j=1}\alpha_{ij}(\mu_j+\phi_j+\xi_j) \right). \label{eq:MU}
\end{equation}
The $\phi_j$ term can be expressed as~\cite{boyerncomp}
\begin{eqnarray}
&& \phi_j=\frac{12}{D}\sum_{\substack{1\leq k<l<m\leq N\\k\neq j, l\neq j, m\neq j}}\Lambda_{j;{j,k,l,m}}C_kC_lC_m.
\label{eq:G}
\end{eqnarray}
We select coefficients $\Lambda_{j;{j,k,l,m}}$ to satisfy the following relation
\begin{dmath}
\sum_{\substack{1\leq j\leq N\\j\neq k,l,m}}\alpha_{ij}\Lambda_{j;{j,k,l,m}}\hiderel=\Gamma_{i,k,l,m},\ \forall i\hiderel\notin{k,l,m},
\\
\Gamma_{i,k,l,m}\hiderel=-6\left(\alpha_{ik}(\sigma_{kl}+\sigma_{km})+\alpha_{il}(\sigma_{kl}+\sigma_{lm})\hiderel+\alpha_{im}(\sigma_{km}+\sigma_{lm})-\gamma_i\right),
\label{eq:ngamma}
\end{dmath}
where $\Gamma_{i,k,l,m}$ is chosen to cancel the $C_kC_lC_m$ terms in Eq.~\ref{eq:fluxwrong} that do not contain $C_i$. The linear system then can be solved to calculate $\Lambda_{j;{j,k,l,m}}$ following
\begin{equation}
(\boldsymbol{\alpha}_{i,j})(\boldsymbol{\Lambda}_j)=(\boldsymbol{\Gamma}_i),\ \forall i,j\hiderel\notin{k,l,m}.
\end{equation}
$(\boldsymbol{\Lambda}_j)$ and $(\boldsymbol{\Gamma}_i)$ are vectors of size $N-3$ with each entry as $\Lambda_{j;{j,k,l,m}}$ and $\Gamma_i$ respectively for all $i,j\notin{k,l,m}$, and $(\boldsymbol{\alpha}_{i.j})$ is the submatrix of $\alpha$ constructed of all entries $i,j$ not equal to $k,l,m$. We can use this to find $\Lambda_{j;{j,k,l,m}}$ for all combinations of $k,l,m$. After following the procedure for $\Lambda_{j;{j,k,l,m}}$, the effect of $\phi_j$ is illustrated in Fig. \ref{fig:ExcessFlux} (b), showing the flux $\mathbf{J}_4 = 0$ when using the modified chemical potential.

In addition, the free energy in Eq.~\ref{eqn:FreeEnergy} is not in general bounded from below outside of the physical range $C_i\in[0,1]$ when one or more spreading parameters
\begin{equation}
\label{eq:spreading}
S_{ij;k}=\sigma_{ij}-\sigma_{ik}-\sigma_{jk}
\end{equation}
are positive~\cite{threecomp,boyerncomp}. We therefore also include an optional stabilising contribution
\begin{eqnarray}
\xi_j=\frac{12}{D}\Omega\sum_{\substack{1\leq k<l\leq N\\k\neq j, l\neq j}}C_k^2C_l^2\left(2C_j-\sum_{\substack{1\leq m\leq N\\m\neq j, k, l}}\Theta_{kl;j;m}C_m\right),\label{eq:P} 
\end{eqnarray}
where we set $\Omega=0$ when all spreading parameters are negative.
The parameters $\Theta_{kl;j;m}$ are evaluated from
\begin{equation}
\sum_{\substack{1\leq j\leq N\\j\neq k,l,m}}\alpha_{ij}\Theta_{kl;j;m}=2\alpha_{im},\ \forall i\hiderel\notin{k,l,m}.  
\end{equation}
This can be equivalently expressed as the linear system
\begin{equation}
(\boldsymbol{\alpha}_{i,j})(\boldsymbol{\Theta}_{k,l})=2(\boldsymbol{\alpha}_{i,m}),\ \forall i\hiderel\notin{k,l,m},
\end{equation}
which we use to obtain all $\Theta_{kl;j;m}$ for a given set of $k,l,m$. This procedure only has to be performed once for a given set of surface tensions and the coefficients can be calculated analytically, so it does not impact the computational performance of the model. We provide a Python script to calculate $\alpha_{ij}$, $\Lambda_{j;{j,k,l,m}}$ and $\Theta_{kl;j;m}$ in the Supplementary Material.

Our strategy to achieve reduction consistency has similarities to that proposed by Boyer and Minjeaud~\cite{boyerncomp}. In their approach, this was done by including an additional contribution to the free energy, and they presented two models. Here, we recast the approach for reduction consistency in terms of contributions to the flux, instead of a global contribution to the free energy. The flux employed in our work, as shown in Eq.~\ref{eq:MU}, maps to the flux that derives from one of their models. In our formulation, it is also unnecessary to include contributions from the $\phi_j$ and $\xi_j$ terms in the chemical force, which is already reduction consistent by default. 
Boyer and Minjeaud~\cite{boyerncomp} further presented an alternative free energy model for arbitrary $N$. However, it is challenging to implement numerically because it is not continuously differentiable with respect to the fluid concentrations.

\section{Lattice Boltzmann Method}

We use the LBM to numerically discretise the continuum model described in Sec. II. In the LBM, the fluid motion is described through the propagation and collision of particle distributions on a regular grid of lattice nodes. The distributions are free to move between lattice sites in a discrete stencil of velocity directions $\mathbf{e}_k$. In this work, we use the D2Q9 stencil, with directions
\begin{equation}
\mathbf{e}=\left[\begin{array}{ccccccccc}
    0 & 1 & -1 & 0 & 0 & 1 & -1 & 1 & -1\\
    0 & 0 & 0 & 1 & -1 & 1 & -1 & -1 & 1
\end{array}\right],
\end{equation}
and weights
\begin{equation}
\mathbf{w}=\left[\begin{array}{ccccccccc}
    \frac{4}{9} & \frac{1}{9} & \frac{1}{9} & \frac{1}{9} & \frac{1}{9} & \frac{1}{36} & \frac{1}{36} & \frac{1}{36} & \frac{1}{36}
\end{array}\right].
\end{equation}
As is standard in the LBM~\cite{lbmbook1}, we set the lattice sound speed as $c_s^2=1/3$ and the time step $\delta t=1$. All quoted quantities are given in lattice units. Extension to three dimensions, such as using a D3Q19 stencil, is straightforward and follows the standard route in the LBM. The computational complexity of our approach with increasing $N$ is discussed in Appendix~\ref{sec:comput}.

\subsection{Hydrodynamics}\label{sec:NmethH}

The lattice Boltzmann equation (LBE) for hydrodynamics can be written as
\begin{align}
f_{k}(\mathbf{x}+\mathbf{e}_k\delta_t,t+\delta_t)-f_{k}(\mathbf{x},t)=\notag\\\Omega_{TRT}(f_k)+F_{TRT}(F_k).
\end{align}
Here, we use the TRT collision operator~\cite{lbmbook1}
\begin{align}
\label{eq:TRT}
\Omega_{TRT}(f_k)=&-\omega^+\delta t(f_k^+(\mathbf{x},t)-f_k^{eq+}(\mathbf{x},t))\notag\\ &-\omega^-\delta t(f_k^-(\mathbf{x},t)-f_k^{eq-}(\mathbf{x},t)),
\end{align}
\begin{equation}
f_k^+=\frac{f_k+f_{\bar{k}}}{2},\qquad f_k^-=\frac{f_k-f_{\bar{k}}}{2},
\end{equation}
where $\bar{k}$ refers to the opposite velocity direction to $k$. $\omega^+$ can be related to the dynamic viscosity in Eq.~\ref{eqn:nse} as
\begin{equation}
\eta=c_s^2\rho\left(\frac{1}{\omega^+\delta t}-\frac{1}{2}\right).
\end{equation}
$\omega^-$ is then a free parameter, and can be used to control the magic parameter $\Lambda^{\textrm{TRT}}$
\begin{equation}
\Lambda^{\textrm{TRT}}=\left(\frac{1}{\omega^+\delta t}-\frac{1}{2}\right)\left(\frac{1}{\omega^-\delta t}-\frac{1}{2}\right),
\end{equation}
which affects the truncation error and stability of the LBM scheme. We use the TRT collision operator to improve accuracy in the case that there are large viscosity contrasts between components~\cite{lbmbook1}. When viscosity ratios are close to $1$, we set $\omega^+=\omega^-$, in which case the TRT collision operator reduces to the simpler BGK approach~\cite{lbmbook1}.

The equilibrium distribution function is defined as~\cite{wellbalanced1,wellbalanced2}
\begin{equation}
f^{eq}_{k}=\begin{cases}
    \frac{P}{c_s^2}(1-w_0)+\rho \left(s_k(\mathbf{v})\right),& \text{if } k=0,\\
    \frac{P}{c_s^2}w_i+\rho \left(s_k(\mathbf{v})\right),& \text{otherwise,}
\label{eq:feq}
\end{cases}
\end{equation}
\begin{equation}
s_k(\mathbf{v})=w_k\left[\frac{\mathbf{v}\cdot\mathbf{e}_k}{c_s^2}+\frac{\left(\mathbf{v}\cdot\mathbf{e}_k\right)^2}{2c_s^4}-\frac{\mathbf{v}\cdot\mathbf{v}}{2c_s^2}\right].
\end{equation}

To incorporate the effects of hydrodynamic and thermodynamic forces, and to enforce incompressibility, we define the forcing term as~\cite{Postma}
\begin{align}
\label{eq:FTRT}
F_{TRT}(F_k)=&\frac{1}{2}\left(1-\frac{1}{2}\omega^+\right)\delta t(F_k+F_{\bar{k}})\notag\\&+\frac{1}{2}\left(1-\frac{1}{2}\omega^-\right)\delta t(F_k-F_{\bar{k}}),
\end{align}
with~\cite{wellbalanced1,wellbalanced2}
\begin{dmath}
\label{eqn:hydroforce}
F_{k}= w_k\left[\frac{\mathbf{v}\cdot\boldsymbol{\nabla}\rho c_s^2+\mathbf{e}_k\cdot \boldsymbol{F}}{c_s^2}+\frac{(\mathbf{v}\cdot\boldsymbol{\nabla}\rho c_s^2+\mathbf{v}\boldsymbol{F}):(\mathbf{e}_k\mathbf{e}_k-c_s^2\boldsymbol{I})}{c_s^4}\right].
\end{dmath}

The surface tension force can be expressed as
\begin{equation}
\label{eq:ststand}
\mathbf{F}_s=\sum_i\mu_i\nabla C_i=\sum_i\left(\frac{\partial E_B}{\partial C_i}+\frac{\partial E_I}{\partial C_i}\right)\nabla C_i,
\end{equation}
which can be equivalently written as the divergence of a stress tensor (proof given in Appendix A)
\begin{equation}
\label{eq:stress_tensor_compact}
\mathbf{F}_s\equiv\nabla\cdot\boldsymbol{\sigma}
= \nabla\cdot\bigg(\left[ E_B + E_I\right]\mathbf{I}
+ \sum_{i\neq j}\kappa_{ij}\,\nabla C_i \otimes \nabla C_j\bigg).
\end{equation}
This equivalence is essential for momentum conservation, as the divergence theorem guarantees that the integral of $\mathbf{F}_s$ over the simulation domain is equal to zero, with solid or periodic boundary conditions. Numerically, however, the equivalence fails because the chain rule,
\begin{equation}
\sum_i \frac{\partial E_B}{\partial C_i}\nabla C_i=\nabla E_B,
\end{equation}
is not satisfied exactly when $\nabla C_i$ and $\nabla E_B$ are calculated through finite difference stencils and $\partial E_B/\partial C_i$ is evaluated analytically. As a result, the standard numerical implementation of $\sum_i\mu_i\nabla C_i$ can no longer be expressed as a divergence, and the divergence theorem no longer applies to guarantee momentum conservation. In Sec. IV A, we demonstrate that the lack of momentum conservation can lead to a large unphysical whole domain drift. To resolve this issue, we modify the chemical force in Eq.~\ref{eq:ststand} with $\sum_i(\partial E_B/\partial C_i)\nabla C_i=\nabla E_B$ to ensure a numerical divergence consistent form
\begin{equation}
\label{eq:Fsnew}
\boldsymbol{F}_s=\nabla{E_B}+\sum_i\frac{\partial E_I}{\partial C_i}\nabla C_i=\nabla{E_B}+\sum_{i,j=1;i\neq j}^N\kappa_{ij}\nabla^2C_j\nabla C_i,
\end{equation}
thereby resolving the domain drift while avoiding the need to evaluate the full stress tensor divergence.

We evaluate all directional gradients using second-order accurate isotropic finite difference stencils
\begin{equation}
\label{eq:gradstencil}
\nabla \psi=\frac{1}{c_s^2\delta t}\sum_{k=1}^{Q-1}w_k\mathbf{e}_k[\psi_{x+\mathbf{e}_k\delta t}-\psi_{x}],
\end{equation}
where $\psi$ is an arbitrary scalar field on the lattice. $\nabla^2 C_i$ needed to calculate $\mu_i$ and $\mathbf{F}_s$ is given by
\begin{equation}
\nabla^2 C_i=\frac{2}{c_s^2\delta t^2}\sum_{k=1}^{Q-1}w_k[C_{i,x+\mathbf{e}_k\delta t}-C_{i,x}].
\end{equation}

The first and zeroth moments of $f_k$ are used to calculate the momentum and hydrodynamic pressure respectively~\cite{wellbalanced1,wellbalanced2}
\begin{dmath}
    \rho\mathbf{v}=\sum_{k=0}^{Q-1}\mathbf{e}_kf_k+\frac{\delta t}{2}\mathbf{F},
\label{eq:nmom}
\end{dmath}
\begin{dmath}
    P=\frac{c_s^2}{1-\omega_0}\left[\sum_{k\neq0}^{Q-1}f_k+\frac{\delta t}{2}\mathbf{v}\cdot\boldsymbol{\nabla}\rho+\rho s_0(\mathbf{v})\right].
\label{eq:pres}
\end{dmath}

\subsection{Interface Capturing}\label{sec:NmethI}

\begin{table*}[!t]
\begin{ruledtabular}
\centering
\caption{Configurations of surface tensions and resulting theoretical and measured Neumann angles in the liquid lens benchmark. The results are plotted in Fig.~\ref{fig:neu} (b). }
\begin{tabular}{cccccc cc cc}
 $\sigma_{12}$ & $\sigma_{13}$ & $\sigma_{14}$ & $\sigma_{23}$ & $\sigma_{24}$ & $\sigma_{34}$ & $\theta_{\mathrm{interior}}^{\mathrm{theory}}$ & $\theta_{\mathrm{exterior}}^{\mathrm{theory}}$ & $\theta_{\mathrm{interior}}^{\mathrm{measured}}$ & $\theta_{\mathrm{exterior}}^{\mathrm{measured}}$\\ 
\midrule$0.005$ & $0.0028$ & $0.0028$ & $0.0028$ & $0.0028$ & $0.0028$& $53.5$ & $153.2$ & $54.2$ & $152.9$  \\
$0.005$ & $0.003$ & $0.003$ & $0.003$ & $0.003$ & $0.003$ & $67.1$ & $146.4$ & $67.4^\dag$ & $146.3^\dag$  \\
$0.005$ & $0.0035$ & $0.0035$ & $0.0035$ & $0.0035$ & $0.0035$ & $88.8$ & $135.6$ & $89.3$ & $135.4$  \\
$0.005$ & $0.004$ & $0.004$ & $0.004$ & $0.004$ & $0.004$ & $102.6$ & $128.7$ & $103.3$ & $128.3$  \\
$0.005$ & $0.005$ & $0.005$ & $0.005$ & $0.005$ & $0.005$ & $120.0$ & $120.0$ & $120.8$ & $119.6$  \\
$0.005$ & $0.006$ & $0.006$ & $0.006$ & $0.006$ & $0.006$ & $130.8$ & $114.6$ & $131.4$ & $114.3$  \\
$0.005$ & $0.007$ & $0.007$ & $0.007$ & $0.007$ & $0.007$ & $138.2$ & $110.9$ & $138.8$ & $110.6$  \\
$0.005$ & $0.008$ & $0.008$ & $0.008$ & $0.008$ & $0.008$ & $143.6$ & $108.2$ & $144.1^\ddag$ & $107.9^\ddag$  \\
$0.005$ & $0.009$ & $0.009$ & $0.009$ & $0.009$ & $0.009$ & $147.7$ & $106.1$ & $148.3$ & $105.9$\\
\end{tabular}
\label{tab:sts}
\end{ruledtabular}
\end{table*}
To evolve the fluid volume fractions $C_i$, we utilise the following LBE~\cite{Zheng2015}
\begin{multline}
g_{i,k}(\mathbf{x}+\mathbf{e}_k\delta_t,t+\delta_t)-g_{i,k}(\mathbf{x},t)=\\-\frac{1}{\tau_{g,i}}\left[g_{i,k}(\mathbf{x},t)-g^{eq}_{i,k}(\mathbf{x},t)\right].
\label{eq:glbe}
\end{multline}
with
\begin{equation}
\label{eq:geq}
g^{eq}_{i,k}=\begin{cases}
    w_kC_i(1+s_k(\mathbf{v}))+\\(1-w_0)m\sum_{j=1}^N\alpha_{ij}(\mu_j+\phi_j+\xi_j),& \text{if } k=0,\\
    w_kC_i(1+s_k(\mathbf{v}))-\\w_km\sum_{j=1}^N\alpha_{ij}(\mu_j+\phi_j+\xi_j),& \text{if } k\neq0.
\end{cases}
\end{equation}
$C_i$ can then be evaluated from the zeroth moment 
\begin{equation}
C_i=\sum_{k=0}^{Q-1} g_{i,k}.
\end{equation}
For diffuse interface models, there are different ways to formulate the fluid viscosity in the interface region. Here, for simplicity, we calculate the fluid viscosity by taking the viscosity of the fluid on each lattice node with the maximum value of $C_i$
\begin{equation}
\eta = \eta_i\ \text{with } i 
\text{ corresponding to } C_i=\max_{j\in\{1,\dots,N\}}C_j.
\end{equation}
The mobility in the Cahn-Hilliard equations can be computed as
\begin{equation}
M=c_s^2m(\tau_{g,i}-0.5)\delta_t.
\end{equation}
In this work, we set $\tau_{g,i}=1$ and $m=1/50$ to fix the mobility as $M=1/300$.

\subsection{Boundary Conditions}

At solid walls, we apply the halfway bounce-back boundary condition~\cite{lbmbook1} for all distributions, which enforces the no-slip condition half a lattice spacing $\frac{1}{2}\mathbf{e}_k\delta t$ beyond the edge of the simulated fluid domain, also implying that the solid wall is at the same position
\begin{equation}
f_{k}^{\mathrm{unknown}}(\mathbf{x}_b,t+\delta t)=f_{\bar{k}}^{*}(\mathbf{x}_b,t+\delta t),
\end{equation}
where $\mathbf{x}_b$ is the position of a node adjacent to the boundary and $f_{\bar{k}}^{*}$ is the updated fluid distribution before propagation
\begin{equation}
f_{\bar{k}}^{*}(\mathbf{x}_b,t+\delta t)=f_{\bar{k}}(\mathbf{x}_{b},t)+\Omega_{TRT}(f_{\bar{k}})+F_{TRT}(F_{\bar{k}}).
\end{equation}
Directional gradients intersecting with solid boundaries are calculated to match the bounce back treatment for the distributions \cite{Lee2010}, where any scalar quantity $\psi_{x+\mathbf{e}_k\delta t}$ in Eq.~\ref{eq:gradstencil} that would lie within a solid wall is replaced with $\psi_{x}$.

In Section~\ref{sec:micro}, we apply a number of additional boundary conditions to allow fluid to enter and exit the simulation domain. The first of these is a Dirichlet condition that enforces a constant velocity on the boundary~\cite{Ladd_1994,lbmbook1}
\begin{equation}
f_{k}^{\mathrm{unknown}}(\mathbf{x}_b,t+\delta t)=f_{\bar{k}}^{*}(\mathbf{x}_b,t+\delta t)-2w_k\rho_b\frac{\mathbf{e}_k\cdot\mathbf{v}_b}{c_s^2},
\label{eq:dirvel}
\end{equation}
where $\rho_b$ and $\mathbf{v}_b$ are the density and velocity at the boundary location respectively.

We also utilise the convective boundary condition~\cite{Convective} to allow fluid to exit the domain according to a convective flow velocity $U$ normal to the boundary following
\begin{equation}
\frac{\partial (\rho\mathbf{v})}{\partial t}+U\frac{\partial (\rho\mathbf{v})}{\partial x}=0,\quad\frac{\partial C_i}{\partial t}+U\frac{\partial C_i}{\partial x}=0,
\end{equation}
which is applied on the boundary location at the edge of the simulated domain. Different choices are available for the value of $U$~\cite{Convective}, but here we take the average normal velocity through the boundary. This boundary condition can be applied to all the distributions on the boundary following
\begin{equation}
f_k(N,t+\delta t) =
\frac{f_k(N,t) + U\frac{\delta t}{\delta x} \; f_k(N-1,t+\delta t)}
{1 + U\frac{\delta t}{\delta x}},
\label{eq:convect}
\end{equation}
where $N$ refers to fluid nodes adjacent to the boundary and $N-1$ refers to nodes adjacent to these further into the fluid domain. Note that here $U$ is evaluated on the $N-1$ nodes after streaming, so we must obtain the velocity on these nodes first, before applying the boundary condition on the nodes at $N$. Directional gradients are evaluated by assuming unknown quantities beyond the boundary $\psi(N+1,t)$ have been convected with velocity $U$\cite{Convective}
\begin{equation}
\psi(N+1,t+\delta t)=\frac{\psi(N+1,t) + U\frac{\delta t}{\delta x} \; \psi(N,t+\delta t)}
{1 + U\frac{\delta t}{\delta x}}.
\end{equation}

Finally, to enforce a constant $C_i$ on the open boundaries, we apply a Dirichlet boundary condition using the anti-bounce-back method~\cite{lbmbook1}, which sets a constant value on the boundary location $\frac{1}{2}\mathbf{e}_k\delta t$ beyond the edge of the simulated nodes
\begin{equation}
g_{k}^{\mathrm{unknown}}(\mathbf{x}_b,t+\delta t)=-g_{\bar{k}}^{*}(\mathbf{x}_b,t+\delta t)+2g^{eq}_{\bar{k}}(C_{i,b},\mathbf{v}_b),
\label{eq:dircomp}
\end{equation}
where $C_{i,b}$ is the volume fraction on the boundary. Directional gradients are evaluated by assuming unknown quantities beyond the boundary match the prescribed Dirichlet value.
On all other boundaries, we apply periodic boundary conditions.

\section{Benchmarks}
\subsection{Liquid Lens and Janus Droplet}
\begin{figure*}[t!]
    \centering
	\includegraphics[width=\textwidth]{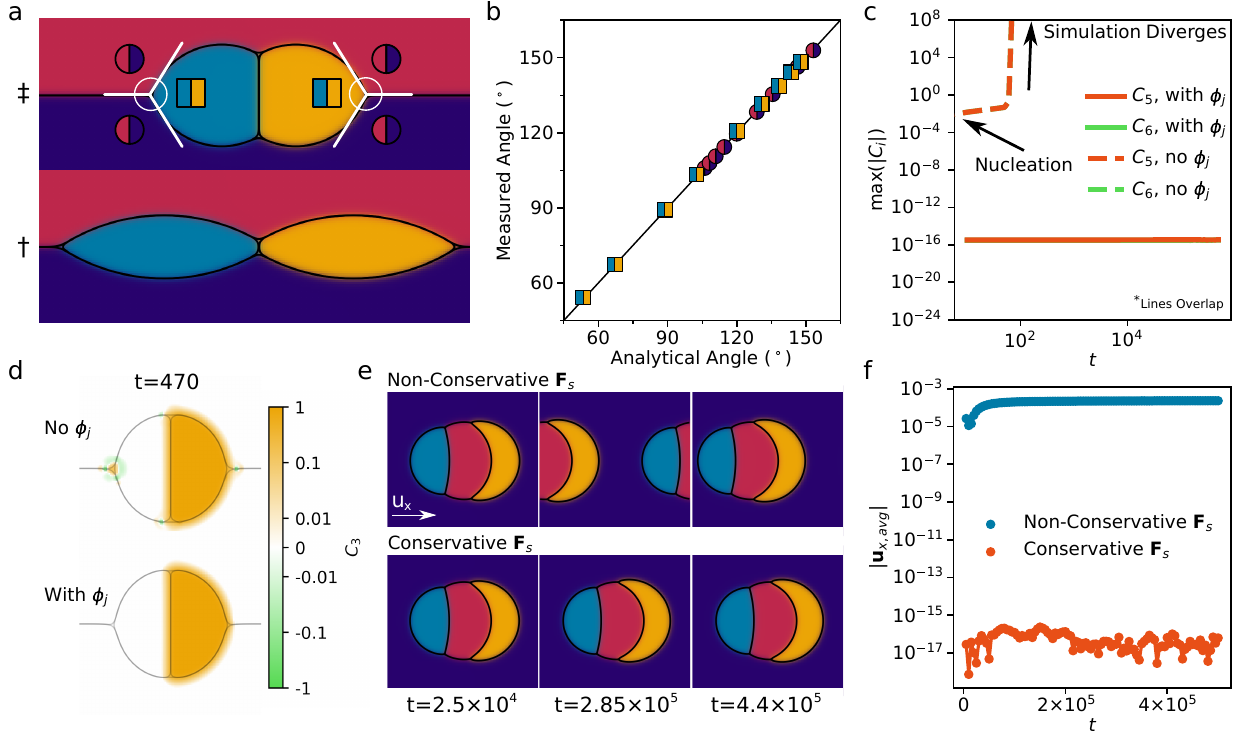}
	\caption{(a) Visualisation of liquid lens setup for configurations $\ddag$ and $\dag$ labelled in Table~\ref{tab:sts}. Symbols associated with the measured interior (square) and exterior (circle) angles between components are shown. Due to symmetry some angles are identical, so we take the average and plot them together. We set $N=6$, and initially set $C_5=C_6=0$ to capture an $N=4$ subsystem. (b) Measured Neumann angles plotted against analytical predictions. (c) Maximum value of $|C_i|$ over time for absent components with the configuration $\dag$, plotted with and without the reduction consistency term $\phi_j$ in the flux. The lines for $C_5$ and $C_6$ overlap. (d) Fluid $C_3$ plotted in a liquid lens setup with $N=4$. We observe nucleation at triple points unless we apply $\phi_j$. (e) Time evolution of an initially stationary triple Janus droplet with the non-conservative surface tension force $\mathbf{F}_s=\sum_i\mu_i\nabla C_i$ and the conservative surface tension force given in Eq.~\ref{eq:Fsnew}. (f) Magnitude of the average velocity of the Janus droplet in the $x$ direction over time.}
	\label{fig:neu}
\end{figure*}
To validate that our method obtains the correct interfacial tensions, we investigate a four fluid liquid-lens setup, with an $N=6$ phase field free energy model to investigate the reduction consistency properties. This involves two droplets at the interface of two other fluids, shown in Fig.~\ref{fig:neu} (a). We set up the simulation by initialising smoothed rectangles of fluid $1$ (purple) and 2 (pink) at the bottom and top halves of a $300\times150$ simulation domain with periodic boundaries. Two smooth semicircles of fluid $3$ (yellow) and $4$ (blue) with a radius of $35$ are initialised in the centre and allowed to evolve towards their equilibrium configuration over $5\times10^5$ timesteps. We set $\eta_i=1/6$ for all fluids.

Theoretical relations for the Neumann angles at three component contact points can be calculated in terms of the surface tensions at each interface triple point~\cite{rowlinson1982molecular}
\begin{equation}
\label{eq:neucosine}
\cos(\pi-\theta_{ij})
= -\frac{\sigma_{ij}^2 - \sigma_{ik}^2 - \sigma_{jk}^2}
        {2 \sigma_{ik}\sigma_{jk}},
\quad 
\substack{i,j,k \in \{1,2,3,4\} \\ i \neq j \neq k},
\end{equation}
where $\theta_{ij}$ refers to the angle formed between the tangents to the interfaces of fluids $i$ and $j$. We group the resulting angles depending on whether they are interior or exterior to the lens fluids, and average the identical angles shown in Fig.~\ref{fig:neu} (a). Angles are obtained fitting the top and bottom of the $C_i=0.5$ contour to two circles and calculating the tangent of these circles at the triple points. We set $\sigma_{34}=0.005$ and vary the remaining independent surface tensions for fluids $1$, $2$, $3$ and $4$ identically to obtain different configurations, shown in Table~\ref{tab:sts}. The surface tensions for absent fluids $5$ and $6$ are all identically $\sigma_{5j}=\sigma_{6j}=0.005$.

Fig.~\ref{fig:neu} (b) and Table~\ref{tab:sts} demonstrate a consistent agreement within $1^\circ$ between theory and measurement. Reduction consistency is essential for this benchmark to guarantee that the correct interfacial tensions are correctly captured in the four component system and that fluids $5$ and $6$ do not nucleate. Fig.~\ref{fig:neu} (c) displays that, if we do not include $\phi_j$, absent fluids nucleate, which quickly drive the system to instability. In contrast, we observe that $C_5$ and $C_6$ remain absent to machine precision when we include $\phi_j$, further confirming that reduction consistency is satisfied.

It should be noted that reduction consistency is necessary even in simulations with no globally absent components. To demonstrate this, we study an identical liquid lens setup with configuration $\dag$ in Fig.~\ref{fig:neu}(a) and Table~\ref{tab:sts} but now set $N=4$, with all fluid components present. Because triple points are effectively local $N=3$ subsystems, we observe that the locally absent $C_3$ nucleates at the triple point formed of fluids $1$, $2$, and $4$ in Fig.~\ref{fig:neu} (d) if we do not resolve reduction consistency by including $\phi_j$ in the flux. In this case, negative values of $C_3$ quickly lead to simulation instability. Including $\phi_j$ completely resolves this issue.

To demonstrate the importance of our conservative surface tension force, we investigate a Janus droplet, consisting of three different fluids. We set $N=4$ and initialise a circular droplet of fluid $2$ with a radius of $20$ in the centre of a $100\times100$ domain with periodic boundaries. We then initialise equivalent droplets of fluids $3$ and $4$ with centres $25$ lattice units to the right and left respectively, minus any overlapping regions. We configure $\sigma_{14}=\sigma_{24}=0.0095$ and set the other surface tensions to $\sigma_{ij}=0.005$. The resulting Janus droplet is asymmetric, as shown in Fig.~\ref{fig:neu} (e), and the non-conservative form of the surface tension force $\mathbf{F}_s=\sum_i\mu_i\nabla C_i$ leads to the droplet drifting over time. This is quantified in Fig.~\ref{fig:neu} (f) by a significant average velocity of $>10^{-4}$ over the whole domain after a short time. Because the largest characteristic velocities in LBM simulations are typically $\leq10^{-2}$ ~\cite{lbmbook1}, this presents a major source of inaccuracy. In contrast, our conservative form effectively eliminates this unphysical drift to machine precision, as shown in Fig.~\ref{fig:neu} (e,f).

\begin{figure}[t!]
    \centering
	\includegraphics[width=\columnwidth]{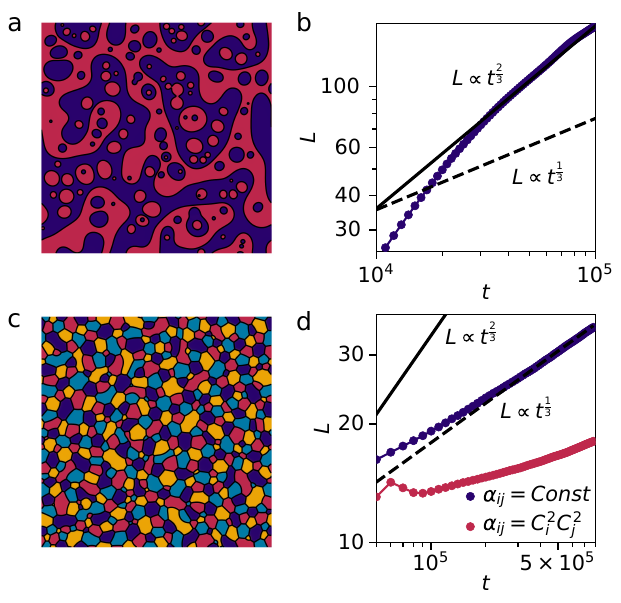}
	\caption{(a) Snapshot during phase separation with $N=6$ and volume fractions $V_1=V_2=1/2$ and $V_3=V_4=V_5=V_6=0$. This case reduces to a binary phase separation problem. (b) The average length scale $L$ from Eq.~\ref{eq:L} follows the expected hydrodynamic $L\propto t^{\frac{2}{3}}$ scaling. (c) Snapshot during phase separation with $V_1=V_2=V_3=V_4=1/4$ and $V_5=V_6=0$ where we do not evolve the fluid velocity. (d) Corresponding data for $L$ over time. Our approach, where $\alpha_{ij}$ are constants obtained from Eqs.~\ref{eq:alpha1} and~\ref{eq:alpha2}, provides the expected $L\propto t^{\frac{1}{3}}$ scaling when compared to the illustrative curve, while enforcing $\alpha_{ij}=C_i^2C_j^2$ fails to capture this regime.}
	\label{fig:phasesep}
\end{figure}
\begin{figure}[t!]
    \centering
	\includegraphics[width=\columnwidth]{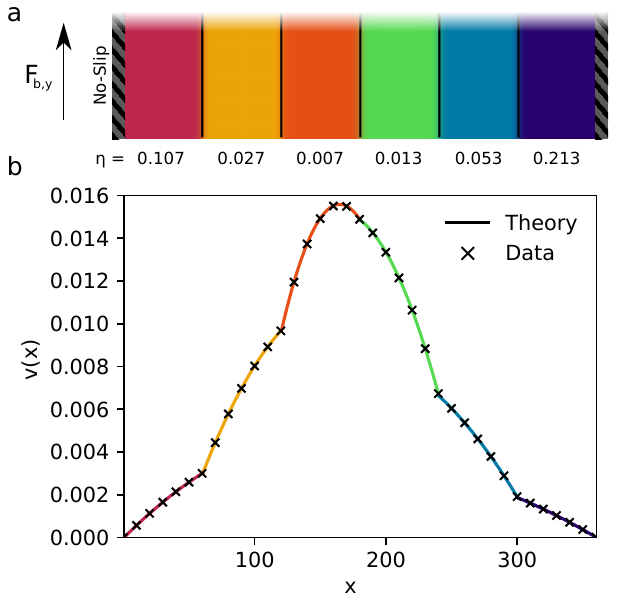}
	\caption{(a) Layered Poiseuille setup, with a body force applied vertically to six fluid layers with different viscosities. We apply the no-slip boundary condition on the left and right walls and periodic boundary conditions on the top and bottom. (b) Equilibrium velocity profile along the channel width plotted against theoretical predictions given by Eq.~\ref{eq:velocityprof}.}
	\label{fig:lay}
\end{figure}

\subsection{Phase Separation}
To apply this model to phase separation problems, we must reproduce established theoretical and experimental domain size scaling predictions. A common measure of the domain size is given by~\cite{phaseseptheo2,phaseseptheo3}
\begin{equation}
L_i=2\pi\frac{\int S_i(k) d^2k}{\int kS_i(k) d^2k},
\label{eq:L}
\end{equation}
which defines a characteristic length scale associated with fluid component $i$. Here, $S_i(k)=\left<\widehat{\,\lambda\,}{}_i(\mathbf{k})\widehat{\,\lambda\,}{}_i(-\mathbf{k})\right>$ is the circularly averaged structure factor, which is obtained from Fourier transforms of the field $\lambda_i(\mathbf{r})=H(C_i(\mathbf{r})-0.5)$, where $H$ is the Heaviside step function. This formulation of $\lambda_i(\mathbf{r})$ is used to avoid contributions from the interface width in $S_i(k)$. We then average $L_i$ across all components to compute the overall domain size $L$. In symmetric two-dimensional binary mixtures, we expect coarsening to occur due to inertial hydrodynamics with $L\propto t^{2/3}$ at late times~\cite{phasesepscaling,phasesepscalingsource,Wagner_2001,WagnerYeomans,Furukawa}. Alternatively, if the primary driver of phase separation is diffusion, the average length scale increases with $L\propto t^{1/3}$ at late times~\cite{phasesepscaling,phasesepscalingsource,lbmphasesep1}. 

We initialise the fluid concentrations by assigning random values of $C_i$ on each lattice site, which are drawn from a uniform distribution in the range $[0, 2/N]$ for each fluid component. These values are then normalized to enforce the constraint $\sum_i C_i = 1$. We set all surface tensions as $\sigma_{ij}=0.005$ and the viscosity of each fluid as $\eta_i=1/6$. In each case, we plot $L$ against $t$ alongside illustrative $L\propto t^{\frac{2}{3}}$ and $L\propto t^{\frac{1}{3}}$ curves. For an $1600\times1600$ $N=2$ system, we observe bicontinuous fluid regions in Fig.~\ref{fig:phasesep} (a) and obtain the expected inertial hydrodynamic scaling of $L\propto t^{\frac{2}{3}}$ in Fig.~\ref{fig:phasesep} (b), with some fluctuations due to finite size effects. If we do not evolve the LBE for Eq.~\ref{eqn:ce} and Eq.~\ref{eqn:nse} and remove the effects of hydrodynamics, we can investigate the diffusive coarsening regime. For $N=4$ and equal compositions in an $400\times400$ system, shown in Fig.~\ref{fig:phasesep} (c), we obtain the correct scaling in Fig.~\ref{fig:phasesep} (d) with $\alpha_{ij}$ evaluated from Eqs.~\ref{eq:alpha1} and~\ref{eq:alpha2}.

Previous approaches~\cite{dongncomp,lbmncomp1,lbmncomp2} have resolved reduction consistency by allowing $\alpha_{ij}$ to vary with the fluid concentrations, such as $\alpha_{ij}=C_iC_j$ or $\alpha_{ij}=C_i^2C_j^2$, with $\alpha_{ii}=-\sum^N_{j=1}\alpha_{ij}$. With these selections, the mobility of an absent component is zero, ensuring it cannot nucleate. However, the first expression for $\alpha_{ij}$ can lead to negative mobilities, causing simulation instability. Moreover, we demonstrate in Fig.~\ref{fig:phasesep} (d) that using $\alpha_{ij}=C_i^2C_j^2$ and $\alpha_{ii}=-\sum^N_{j=1}\alpha_{ij}$ cannot reproduce the $L\propto t^{\frac{1}{3}}$ scaling, as diffusion is now concentrated around the fluid interface. Therefore, previous approaches to capture reduction consistency in the LBM are unsuitable for complete study of phase separation dynamics. The procedure to incorporate $\alpha_{ij}=C_i^2C_j^2$ in the LBM is described in Appendix C.

\subsection{Layered Poiseuille Flow}

In order to quantitatively validate the dynamic flow performance of the method, we investigate layered Poiseuille flow. The setup is shown in Fig. 3(a), where six fluid layers are introduced into an $N=6$ system parallel to two solid boundaries, with a domain size of $360\times1$. We apply periodic boundaries on the top and bottom of the domain, and apply a constant body force parallel to the solid walls $F_{b,y}=4\times10^{-8}$. We set $\omega^+=1/0.82$, $1/0.58$, $1/0.52$, $1/0.54$, $1/0.66$ and $1/1.14$ and $\rho=1$ to obtain $\eta=0.107$, $0.027$, $0.007$, $0.013$, $0.053$ and $0.213$ respectively, and set $\Lambda^{\textrm{TRT}}=3/16$. The surface tensions are all $\sigma_{ij}=0.005$.
\begin{figure*}[t!]
    \centering
	\includegraphics[width=\textwidth]{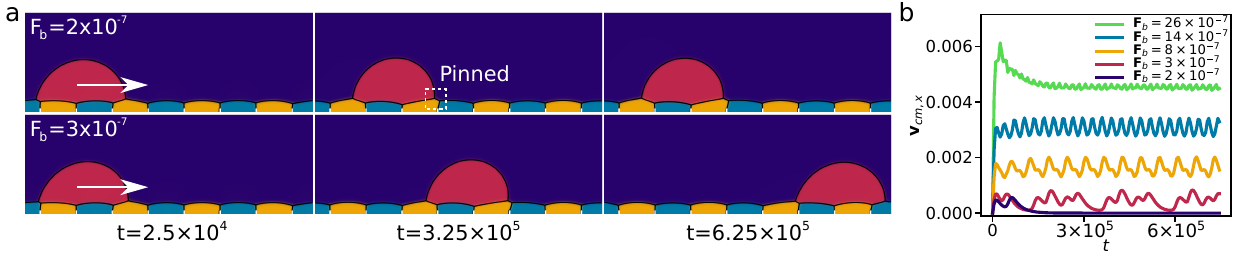}
	\caption{(a) Visualisation of PaLS setup, with a droplet of fluid $2$ (pink) on alternating patches of fluid $3$ (yellow) and fluid $4$ (blue) surrounded by fluid $1$ (purple). Droplets on PaLS can become pinned on lubricant-lubricant-air contact lines~\cite{xitong}.
    Applying a large enough body force $\mathbf{F}_b$ allows the droplet to overcome the pinning force and subsequently travel over the surface. (b) Centre of mass velocity of droplets moving on PaLS, with varying $\mathbf{F}_b$.}
	\label{fig:pals}
\end{figure*}

To derive an analytical solution for the equilibrium velocity profile, we note that the viscous force and the applied body force exactly cancel in equilibrium
\begin{equation}
\eta_i \frac{d^2 v_i}{d x^2} = F_{b,y},
\end{equation}
for each layer $i$. Here, $y$ is the flow direction and $x$ is normal to the left solid wall. We can integrate once to obtain
\begin{equation}
\frac{d v_i}{d x} = \frac{F_{b,y}}{\eta_i} x + A_i,
\label{eq:piecewisequad}
\end{equation}
where $A_i$ is an integration constant. Applying the constraint of continuous shear stress across the interface
\begin{equation}
\eta_i \frac{dv_i}{dx}(x_i)= \eta_{i+1} \frac{dv_{i+1}}{dx}(x_i),
\end{equation}
where $x_i$ is the location of the interface between layers $i$ and $i+1$, we conclude that $\eta_iA_i=A$ is an identical constant for all layers. Integrating again, we arrive at the velocity profile
\begin{equation}
v_i(x) = \frac{F_{b,y}}{2\eta_i} x^2 + \frac{A}{\eta_i} x + B_i,
\label{eq:velocityprof}
\end{equation}
where $B_i$ is an integration constant for each layer. Now, we use the fact that the velocity is continuous across each interface, and apply the no-slip boundary condition at the solid walls
\begin{align}
v_i(x_i) &= v_{i+1}(x_i),\quad v_1(0) = 0,\quad v_6(W) = 0.
\end{align}
The result is a linear system that can be solved to obtain the
constants $A$ and $B_i$
\begin{equation}
\mathbf{A}\mathbf{x} = \mathbf{b},
\quad 
\mathbf{x} =
(A,B_1,B_2,B_3,B_4,B_5,B_6)^T.
\end{equation}
From this system of equations, we obtain $A=-6.57\times10^{-6}$, $B_1=0$, $B_2=0.00906$, $B_3=0.0654$, $B_4=0.0253$, $B_5=0.00135$ and $B_6=-0.00106$.

In Fig.~\ref{fig:lay}(b), we show the measured velocity profile compared with the theoretical curves given by Eq.~\ref{eq:velocityprof}. We obtain an excellent agreement between the simulated data and the theoretical solution for the wide range of viscosities tested.

\section{Applications}

\subsection{Patterned Liquid Surfaces}

Patterned liquid surfaces (PaLS) are solid surfaces coated with thin domains of two immiscible lubricant phases arranged in prescribed patterns~\cite{pal1,pal2}. In contrast to traditional liquid-infused surfaces, which exploit a smooth lubricant coating to repel droplets at tilt angles of $\approx1-3^\circ$~\cite{lisangle1,lisangle2}, PaLS offer the possibility to direct droplet motion by exploiting pinning on the interfaces between lubricant patches. Recent experimental and theoretical studies have revealed a rich phase space of distinct wetting states of droplets on these surfaces~\cite{xitong}, but computational approaches so far have been limited to static energy-minimisation simulations. In this section, we highlight the applicability of our approach to study these systems dynamically for the first time. 

We initialise eight stripes of alternating lubricant components of fluids $3$ and $4$ with a height of 15 in a $400\times140$ simulation domain, with solid walls on the top and bottom and periodic boundaries on the sides. We then initialise a hemispherical droplet of fluid $2$ with radius of 60 surrounded by fluid $1$. To prevent the lubricant phases from travelling over the surface, we place solid posts with dimensions $2\times5$ on the surface between each lubricant-lubricant interface. In experiments, the lubricant is prevented from moving by chemically patterning the underlying substrate~\cite{xitong}. To investigate droplet motion on these surfaces, we apply a horizontal body force to the droplet of fluid $2$ and vary the magnitude. The surface tensions are configured as $\sigma_{13}=\sigma_{14}=\sigma_{23}=\sigma_{24}=0.01$ and $\sigma_{12}=\sigma_{34}=0.005$, with $\Omega=1/90$ in Eq.~\ref{eq:P}. The viscosities are $\eta_1=2/300$, $\eta_2=0.1$ and $\eta_3=\eta_4=2/3$ and we set $\Lambda^{\textrm{TRT}}=3/16$.

With a body force of $\mathbf{F}_b=2\times10^{-7}$, the front of the droplet becomes pinned at the first lubricant-lubricant-air contact line it approaches, shown in Fig.~\ref{fig:pals} (a). Increasing the body force to $\mathbf{F}_b=3\times10^{-7}$ allows the droplet to overcome the pinning force and it begins to move over the surface. The full time evolution in each case is shown in Supplementary Movie 1 and 2 respectively. We plot the droplet centre-of-mass velocity as $\mathbf{v}_{cm,x}=\int C_2\mathbf{v}_xdV/\int C_2dV$ for varying $\mathbf{F}_b$ in Fig.~\ref{fig:pals} (b). At smaller $\mathbf{F}_b$, the droplet exhibits stick-slip motion, as it easily slides over the smooth lubricant interfaces, before it is slowed at the lubricant-lubricant-air contact lines. As a result, the centre of mass velocity oscillates over time. Because the viscous friction force scales with the droplet velocity~\cite{Naga2024,keiser_drop_2017,daniel_oleoplaning_2017}, at higher body forces, the magnitude of the oscillations reduces as viscous drag becomes the dominant source of friction.
\begin{figure}[t!]
    \centering
	\includegraphics[width=\columnwidth]{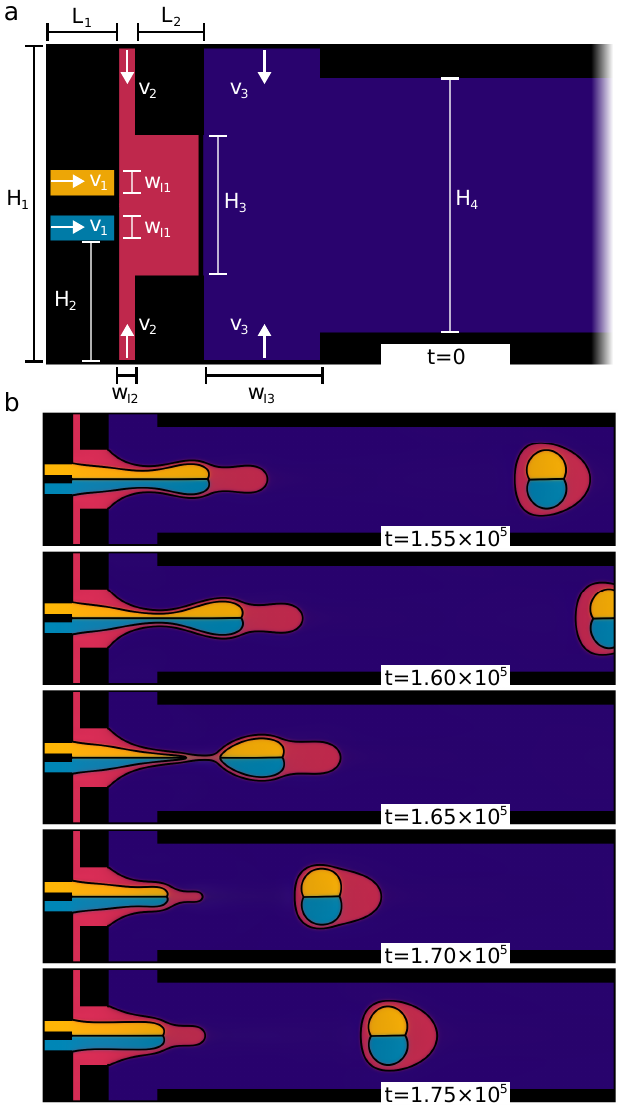}
	\caption{(a) Microfluidic design for generating triple-emulsion droplets, showing the initial state at $t=0$. Physical dimensions of the design are annotated and the arrows denote flow directions from the inlets. The visualisation is cut off before the edge of the simulation domain on the right hand side. (b) Time evolution showing the formation of an emulsion droplet from the injected fluids. The surface tensions are configured to encourage cloaking of fluids $3$ (yellow) and $4$ (blue) by fluid $2$ (pink).}
	\label{fig:microfluidic}
\end{figure}
\subsection{Emulsion Droplet Generation}\label{sec:micro}

Emulsion droplets are hierarchical fluid structures comprising a host droplet with internal droplets of other immiscible phases. The generation of emulsion droplets is commonly achieved using microfluidic devices~\cite{microshear,microemulsion1,Janusmicrofluidic,microemulsion2,microemulsion3,microemulsion4,microfluidic5}, and the resulting structures can be employed to facilitate drug encapsulation and delivery~\cite{microemulsion4}, enable controlled chemical reactions~\cite{microreactor2,microreactor1}, and act as templates for Janus particle fabrication~\cite{Janusmicrofluidic,microemulsion3}. The inclusion of $\xi_j$ in Eq.~\ref{eq:P} enables our method to study the rich phase space of possible distinct droplet morphologies which can form with different combinations of positive spreading parameters $S_{ij;k}>0$.

To study the generation of emulsion droplets, we employ a microfluidic geometry, shown in Fig.~\ref{fig:microfluidic} (a), designed to sequentially introduce and encapsulate multiple immiscible fluid phases.
The geometry consists of two inlets with length $L_1=30$ and width $w_{I1}=10$ on the left of the domain, one of which injects pure fluid $3$ and the other injects pure fluid $4$, both with velocity $v_1=0.003$. The total domain height is $H_1=140$ and each channel begins a distance of $H_2=55$ from the edge of the domain. These are followed by two perpendicular inlets of fluid $2$ with width $w_{I2}=8$ at velocity $v_2=0.0045$. The fluids then enter into a wider channel with height $H_3=80$ and length $L_2=30$ followed by two further inlets for fluid $1$ at velocity $v_3=0.0098$ and width $w_{I3}=51$. Finally, the geometry opens into a channel with height $H_4=110$. We configure $\sigma_{14}=\sigma_{13}=0.012$ and the remaining surface tensions as $\sigma_{ij}=0.005$ to encourage fluid $2$ to completely cloak fluids $3$ and $4$, and we set $\Omega=1/18$ in Eq.~\ref{eq:P}. The viscosities are identically $\eta_i=1/12$. At the inlets, we apply Dirichlet conditions for the velocity and volume fractions $C_i$ using Eq.~\ref{eq:dirvel} and Eq.~\ref{eq:dircomp} respectively.
The full domain size is $600\times140$, and we apply the convective boundary condition in Eq.~\ref{eq:convect} on the right boundary.

Fig.~\ref{fig:microfluidic} (b) shows snapshots of the droplet formation process over time. As fluids $3$ and $4$ are injected into fluid $2$, a composite structure comprising all three fluids is formed and flows into the wider downstream channel, where it is elongated into a compound thread by the flow of fluid $1$. As the thread extends, interfacial tension acts to minimize the surface area, ultimately leading to capillary instability and pinch-off, resulting in the formation of an emulsion droplet of fluid $2$ that encapsulates one droplet each of fluids $3$ and $4$. The design continues to produce regular emulsion droplets over time, as evidenced by the previously generated droplet visible in the first two snapshots. The full time evolution is shown in Supplementary Movie 3.

\section{Conclusion}

In this paper, we have presented a free energy lattice Boltzmann method (LBM) for simulating flows with an arbitrary number of immiscible fluid components in principle. The method allows for independent selection of the interfacial tensions and viscosities
for each fluid component in the system. Our underlying model is strictly reduction consistent, ensuring that an $N$ component system exactly recovers the corresponding $N-1$ system when one component is absent. A key advantage of our approach is that reduction consistency is enforced independently of the diffusive mobility, enabling accurate control of diffusive dynamics and the recovery of established phase-separation scaling laws. In addition, we introduced a momentum conserving discretisation of the surface tension force, resolving a significant velocity drift present in previous LBM schemes.

We demonstrate the capabilities of our method using a series of static and dynamic benchmark problems, including liquid lenses, Janus droplets, layered Poiseuille flow and liquid-liquid phase separation, and obtain excellent agreement throughout between theoretical and simulated results. We further highlight the promising capabilities of the method for future study of applications involving three or more immiscible fluids, including droplet dynamics on patterned liquid infused surfaces and microfluidic devices for multiple emulsion droplet generation.

Our method is well positioned to support existing experimental work in a broad range of applications involving multiple immiscible fluids. Potential areas include biomedical systems, such as the study of phase-separated biomolecular condensates~\cite{phasesepcondensate,zwicker,phasesepmany1,phasesepmany2} and droplet-based cell assays~\cite{Lareau2019,Zheng2017}; new soft materials, including DNA nanostars~\cite{chaderjian2025diversedistinctdenselypacked} and block copolymers~\cite{blockcopolymer,Lang2022,Peinemann2007}; patterned liquid surfaces~\cite{xitong,pal1,pal2} to direct droplet motion; complex emulsions and layered textures in food~\cite{KLOJDOVA2022100451} and cosmetic~\cite{cosmetics3040037} formulations; and microreactors for controlled chemical reactions~\cite{microreactor2,microreactor1}. Future extensions to the method should include wetting interactions with solid boundaries~\cite{dongwall} and support for large density contrasts among the fluid components~\cite{Lee2010,PhysRevE.97.033312}.

\section*{Supplementary Material}

See the Supplementary Material for supplementary movies associated with Fig. 5 (a) and Fig. 6 (b), and a Python script to evaluate $\alpha_{ij}$, $\Lambda_{j;j,k,l,m}$, and $\Theta_{kl;j;m}$ in Sec.~\ref{sec:continuum}.

\begin{acknowledgments}
We thank Dr Rodrigo Ledesma and Dr Luiz Eduardo Czelusniak for useful discussions. This work was supported by Leverhulme Trust (research project grant RPG-2022-140), UKRI Engineering and Physical Sciences Research Council (EP/V034154/2), and EPSRC IAA project ``Liquid Evaporation of Structured Surface''. This work used the Cirrus UK National Tier-2 HPC Service at EPCC. We also acknowledge compute resources on ARCHER2 via the UK Consortium on Mesoscale Engineering Sciences (EP/L00030X/1). In addition, we want to thank the DFG project OPUS-26 (LAP). For the purpose of open access, the author has applied a Creative Commons Attribution (CC BY) licence to any Author Accepted Manuscript version arising from this submission.
\end{acknowledgments}

\section*{Conflict of Interest Statement}

The authors have no conflicts to disclose.

\section*{Author Contributions}

\textbf{Michael Rennick:} Conceptualization (equal); Investigation (lead); Methodology (lead); Software (lead); Writing - original draft (equal); Writing - review \& editing (equal). \textbf{Xitong Zhang:}  Conceptualization (equal); Methodology (supporting); Writing - review \& editing (equal). \textbf{Tim Niklas Bingert:} Methodology (supporting); Software (supporting); Writing - review \& editing (equal). \textbf{Mathias J. Krause:} Funding acquisition (equal); Supervision (supporting); Writing - review \& editing (equal). \textbf{Halim Kusumaatmaja:} Conceptualization (equal); Funding acquisition (equal); Supervision (lead); Writing - original draft (equal); Writing - review \& editing (equal).

\section*{Data Availability Statement}

The data that support the findings of this study are available from the corresponding author upon reasonable request.

\appendix

\section{Equivalence of Chemical Force and Stress Tensor Divergence}
The divergence of the stress tensor $\nabla\cdot\boldsymbol{\sigma}$ is analytically consistent with the application of a chemical force $\sum_i\mu_i\nabla C_i$. The stress tensor $\boldsymbol{\sigma}\equiv\sigma_{\alpha\beta}$ can be written in tensor notation as
\begin{align}
\label{eq:Nstress}
\sigma_{\alpha\beta}
=& \delta_{\alpha\beta}\!\left[E_B-\sum_{i,j=1;i\neq j}\frac{\kappa_{ij}}{2}(\partial_\gamma C_i)(\partial_\gamma C_j)
\right]\notag
\\&+ \sum_{i,j=1;i\neq j}^N\kappa_{ij} \, (\partial_\alpha C_i)(\partial_\beta C_j).
\end{align}
The divergence is then given by $\nabla\cdot\boldsymbol{\sigma}\equiv\partial_\beta\sigma_{\alpha\beta}$. For the second term in square brackets, we apply the chain rule
\begin{equation}
\partial_\beta(\delta_{\alpha\beta}E_B)=\sum_i \frac{\partial E_B}{\partial C_i}\partial_\alpha C_i.
\end{equation}
For the third and fourth terms, we use the product rule
\begin{align}
\partial_\beta\bigg(\delta_{\alpha\beta}\sum_{i,j=1;i\neq j}&\frac{\kappa_{ij}}{2}(\partial_\gamma C_i)(\partial_\gamma C_j)\bigg)=
 \notag\\ \sum_{i,j=1;i\neq j}\frac{\kappa_{ij}}{2}&\left[(\partial_\alpha\partial_\gamma C_i)(\partial_\gamma C_j)
+(\partial_\gamma C_i)(\partial_\alpha\partial_\gamma C_j)\right],
\end{align}
\begin{align}
\partial_\beta\bigg(\delta_{\alpha\beta}\sum_{i,j=1;i\neq j}&\kappa_{ij}(\partial_\alpha C_i)(\partial_\beta C_j)\bigg)=\notag\\ \sum_{i,j=1;i\neq j}\kappa_{ij}&\left[(\partial_\beta\partial_\alpha C_i)(\partial_\beta C_i))+(\partial_\alpha C_i)(\partial_\beta\partial_\beta C_i))\right].
\end{align}
Subsequently, we can express the total divergence as
\begin{dmath}
\label{eq:divkorteweg1}
\partial_\beta \sigma_{\alpha\beta}=\sum_i \frac{\partial E_B}{\partial C_i}\partial_\alpha C_i-\sum_{i,j=1;i\neq j}\frac{\kappa_{ij}}{2}\left[(\partial_\alpha\partial_\gamma C_i)(\partial_\gamma C_j)+(\partial_\gamma C_i)(\partial_\alpha\partial_\gamma C_j)\right]+\sum_{i,j=1;i\neq j}\kappa_{ij}\left[(\partial_\beta\partial_\alpha C_i)(\partial_\beta C_j))+(\partial_\alpha C_i)(\partial_\beta\partial_\beta C_j))\right].
\end{dmath}
The chemical potential can be expressed in terms of bulk and gradient terms as
\begin{equation}
\mu_i=\frac{\partial E_B}{\partial C_i}+\sum_{j=1,i\neq j}\kappa_{ij}\partial_\gamma\partial_\gamma C_j.
\end{equation}
We can multiply this by $\partial_\alpha C_i$ to obtain
\begin{equation}
\label{eq:mugradphi}
\mu_i\partial_\alpha C_i=\frac{\partial E_B}{\partial C_i}\partial_\alpha C_i+\sum_{j=1,i\neq j}\kappa_{ij}(\partial_\gamma\partial_\gamma C_j)(\partial_\alpha C_i),
\end{equation}
\begin{equation}
\label{eq:subs}
\sum_i\frac{\partial E_B}{\partial C_i}\partial_\alpha C_i=\sum_i\mu_i\partial_\alpha C_i-\sum_{i,j=1,i\neq j}\kappa_{ij}(\partial_\gamma\partial_\gamma C_j)(\partial_\alpha C_i).
\end{equation}
After substituting Eq~\ref{eq:subs} into Eq.~\ref{eq:divkorteweg1}, we arrive at
\begin{dmath}
\partial_\beta \sigma_{\alpha\beta}=\sum_i\mu_i\partial_\alpha C_i-\sum_{i,j=1,i\neq j}\kappa_{ij}(\partial_\gamma\partial_\gamma C_j)(\partial_\alpha C_i)-\sum_{i,j=1;i\neq j}\frac{\kappa_{ij}}{2}\left[(\partial_\alpha\partial_\gamma C_i)(\partial_\gamma C_j)+(\partial_\gamma C_i)(\partial_\alpha\partial_\gamma C_j)\right]+\sum_{i,j=1;i\neq j}\kappa_{ij}\left[(\partial_\beta\partial_\alpha C_i)(\partial_\beta C_j))+(\partial_\alpha C_i)(\partial_\beta\partial_\beta C_j))\right].
\end{dmath}
After exploiting the symmetry $\kappa_{ij}=\kappa_{ji}$ and relabelling indices, the mixed gradient terms in $C_i$ cancel and we are finally left with
\begin{equation}
\label{eq:kortewegequiv}
\partial_\beta \sigma_{\alpha\beta}=\sum_i\mu_i\partial_\alpha C_i.
\end{equation}

\section{Computational Complexity}\label{sec:comput}

We expect the lattice Boltzmann equations for the evolution of $f_k$ and $g_k$ to occupy the majority of the computational time at small $N$. For each added phase, we must solve an additional lattice Boltzmann equation and the complexity increases with $\mathcal{O}(N)$. Furthermore, the complexity of the chemical potential and stress tensor gradient calculation scales with $\mathcal{O}(N^2)$ and the calculation of $\phi_j$ and $\xi_j$ scales with $\mathcal{O}(N^3)$, and these calculations will start to dominate at very large $N$. The developed method in this work is best suited to fluid problems with $N\lesssim\mathcal{O}(10)$, where independent selection of some or all interfacial tensions is necessary. For example, this is relevant to model recent experimental systems using DNA nanostars with up to 9 immiscible phases~\cite{chaderjian2025diversedistinctdenselypacked}. In the future, it is also possible to consider incorporating a near contact force~\cite{TIRIBOCCHI20251,Montessori_Lauricella_Tirelli_Succi_2019} in our method. This would allow for independent selection of some interfacial tensions while improving computational complexity for large numbers of droplets.

\section{Implementation of Concentration Dependent Mobility}

In Fig.~\ref{fig:phasesep} (d), implementation of the concentration dependent mobility
\begin{equation}
\label{eq:mobilconc}
\alpha_{ij}=C_i^2C_j^2,\qquad \alpha_{ii}=-\sum^N_{j=1}\alpha_{ij},
\end{equation}
is performed following the approach of Zheng et al. (2021)~\cite{lbmncomp1}. The difference in approach is necessary to avoid spatial gradients in $\alpha_{ij}$ in the diffusive flux. Here, we set $m=0$ in Eq.~\ref{eq:geq} and instead implement the right hand side of Eq.~\ref{eq:CHFlux} by applying a source term to the LBE for distributions $g_{i,k}$
\begin{align}
&g_{i,k}(\mathbf{x}+\mathbf{e}_k\delta_t,t+\delta_t)-g_{i,k}(\mathbf{x},t)=\notag\\&-\frac{1}{\tau_{g,i}}\left[g_{i,k}(\mathbf{x},t)-g^{eq}_{i,k}(\mathbf{x},t)\right]+\delta t\bigg(1-\frac{1}{2\tau_{g,i}}\bigg)w_km^*\mathbf{e}_k\cdot\mathbf{G}_i,
\label{eq:glbesource}
\end{align}
\begin{equation}
\mathbf{G}_i= \sum^N_{j=1}\alpha_{ij}\mathbf{\nabla}(\mu_j+\phi_j+\xi_j).
\end{equation}
The mobility is given by
\begin{equation}
M=\delta_tc_s^2m^*(\tau_{g,i}-0.5),
\end{equation}
and we set $m^*=10$.

\bibliography{refs}

@PREAMBLE{
 "\providecommand{\noopsort}[1]{}" 
 # "\providecommand{\singleletter}[1]{#1}%" 
}

@article{phasesepscalingsource,
author = {Hiroshi Furukawa},
title = {A dynamic scaling assumption for phase separation},
journal = {Advances in Physics},
volume = {34},
number = {6},
pages = {703--750},
year = {1985},
publisher = {Taylor \& Francis},
doi = {10.1080/00018738500101841},


URL = { 
    
        https://doi.org/10.1080/00018738500101841
    
    

}
}

@article{phasesepscaling,
  title = {Breakdown of Dynamic Scaling in Thin Film Binary Liquids Undergoing Phase Separation},
  author = {Chung, Hyun-Joong and Composto, Russell J.},
  journal = {Physical Review Letters},
  volume = {92},
  issue = {18},
  pages = {185704},
  numpages = {4},
  year = {2004},
  month = {May},
  publisher = {American Physical Society},
  doi = {10.1103/PhysRevLett.92.185704}
}

@Article{phaseseptheo2,
author ="Mao, Sheng and Kuldinow, Derek and Haataja, Mikko P. and Košmrlj, Andrej",
title  ="Phase behavior and morphology of multicomponent liquid mixtures",
journal  ="Soft Matter",
year  ="2019",
volume  ="15",
issue  ="6",
pages  ="1297-1311",
publisher  ="The Royal Society of Chemistry",
doi  ="10.1039/C8SM02045K"}

@article{phaseseptheo3, title={Inertial effects in three-dimensional spinodal decomposition of a symmetric binary fluid mixture: a lattice Boltzmann study}, volume={440}, DOI={10.1017/S0022112001004682}, journal={Journal of Fluid Mechanics}, author={Kendon, Vivien M. and Cates, Michael E. and Pagonabarraga, Ignacio and Desplat, J.-C. and Bladon, Peter}, year={2001}, pages={147–203}}

@article{phasesepmany1,
    author = {Jacobs, William M. and Frenkel, Daan},
    title = {Predicting phase behavior in multicomponent mixtures},
    journal = {The Journal of Chemical Physics},
    volume = {139},
    number = {2},
    pages = {024108},
    year = {2013},
    month = {07},
    issn = {0021-9606},
    doi = {10.1063/1.4812461}
}

@Article{phasesepmany2,
author={Jacobs, William M.
and Frenkel, Daan},
title={Phase Transitions in Biological Systems with Many Components},
journal={Biophysical Journal},
year={2017},
month={Feb},
day={28},
publisher={Elsevier},
volume={112},
number={4},
pages={683-691},
issn={0006-3495},
doi={10.1016/j.bpj.2016.10.043}
}

@article{
phasesepcondensate,
author = {Yongdae Shin  and Clifford P. Brangwynne },
title = {Liquid phase condensation in cell physiology and disease},
journal = {Science},
volume = {357},
number = {6357},
pages = {eaaf4382},
year = {2017},
doi = {10.1126/science.aaf4382}}

@Article{microemulsion1,
author ="Wang, Wei and Xie, Rui and Ju, Xiao-Jie and Luo, Tao and Liu, Li and Weitz, David A. and Chu, Liang-Yin",
title  ="Controllable microfluidic production of multicomponent multiple emulsions",
journal  ="Lab Chip",
year  ="2011",
volume  ="11",
issue  ="9",
pages  ="1587-1592",
publisher  ="The Royal Society of Chemistry",
doi  ="10.1039/C1LC20065H"}

@ARTICLE{microemulsion2,
  title    = "Dielectrophoresis Response of {Water-in-Oil-in-Water} Double
              Emulsion Droplets with Singular or Dual Cores",
  author   = "Jiang, Tianyi and Jia, Yankai and Sun, Haizhen and Deng, Xiaokang
              and Tang, Dewei and Ren, Yukun",
  journal  = "Micromachines (Basel)",
  volume   =  11,
  number   =  12,
  month    =  dec,
  year     =  2020,
  address  = "Switzerland"
}

@article{microemulsion3,
title = {Microfluidic emulsification techniques for controllable emulsion production and functional microparticle synthesis},
journal = {Chemical Engineering Journal},
volume = {452},
pages = {139277},
year = {2023},
issn = {1385-8947},
doi = {https://doi.org/10.1016/j.cej.2022.139277},
author = {Wei Wang and Bing-Yu Li and Mao-Jie Zhang and Yao-Yao Su and Da-Wei Pan and Zhuang Liu and Xiao-Jie Ju and Rui Xie and Yousef Faraj and Liang-Yin Chu}
}

@article{
microemulsion4,
author = {Jing Zhang  and Roger J. Coulston  and Samuel T. Jones  and Jin Geng  and Oren A. Scherman  and Chris Abell },
title = {One-Step Fabrication of Supramolecular Microcapsules from Microfluidic Droplets},
journal = {Science},
volume = {335},
number = {6069},
pages = {690-694},
year = {2012},
doi = {10.1126/science.1215416},
URL = {https://www.science.org/doi/abs/10.1126/science.1215416}}

@Article{pal1,
author={Pelizzari, Michele
and McHale, Glen
and Armstrong, Steven
and Zhao, Hongyu
and Ledesma-Aguilar, Rodrigo
and Wells, Gary G.
and Kusumaatmaja, Halim},
title={Droplet Self-Propulsion on Slippery Liquid-Infused Surfaces with Dual-Lubricant Wedge-Shaped Wettability Patterns},
journal={Langmuir},
year={2023},
month={Nov},
day={07},
publisher={American Chemical Society},
volume={39},
number={44},
pages={15676-15689},
issn={0743-7463},
doi={10.1021/acs.langmuir.3c02205},
}

@Article{pal2,
author={Paulssen, Dorothea
and Hardt, Steffen
and Levkin, Pavel A.},
title={Droplet Sorting and Manipulation on Patterned Two-Phase Slippery Lubricant-Infused Surface},
journal={ACS Applied Materials {\&} Interfaces},
year={2019},
month={May},
day={01},
publisher={American Chemical Society},
volume={11},
number={17},
pages={16130-16138},
issn={1944-8244},
doi={10.1021/acsami.8b21879},
}

@article{lbmapplication1,
  title = {Convection in Multiphase Fluid Flows Using Lattice Boltzmann Methods},
  author = {Biferale, L. and Perlekar, P. and Sbragaglia, M. and Toschi, F.},
  journal = {Physical Review Letters},
  volume = {108},
  issue = {10},
  pages = {104502},
  numpages = {5},
  year = {2012},
  month = {Mar},
  publisher = {American Physical Society},
  doi = {10.1103/PhysRevLett.108.104502},
}

@Article{lbmapplication2,
author={Falcucci, Giacomo
and Amati, Giorgio
and Fanelli, Pierluigi
and Krastev, Vesselin K.
and Polverino, Giovanni
and Porfiri, Maurizio
and Succi, Sauro},
title={Extreme flow simulations reveal skeletal adaptations of deep-sea sponges},
journal={Nature},
year={2021},
month={Jul},
day={01},
volume={595},
number={7868},
pages={537-541},
issn={1476-4687},
doi={10.1038/s41586-021-03658-1}
}

@article{lbmapplication3,
  title = {Ternary Free-Energy Entropic Lattice Boltzmann Model with a High Density Ratio},
  author = {W\"ohrwag, M. and Semprebon, C. and Mazloomi Moqaddam, A. and Karlin, I. and Kusumaatmaja, H.},
  journal = {Physical Review Letters},
  volume = {120},
  issue = {23},
  pages = {234501},
  numpages = {6},
  year = {2018},
  month = {Jun},
  publisher = {American Physical Society},
  doi = {10.1103/PhysRevLett.120.234501}
}

@Article{lbmphasesep1,
author ="Shek, Alvin C. M. and Kusumaatmaja, Halim",
title  ="Spontaneous phase separation of ternary fluid mixtures",
journal  ="Soft Matter",
year  ="2022",
volume  ="18",
issue  ="31",
pages  ="5807-5814",
publisher  ="The Royal Society of Chemistry",
doi  ="10.1039/D2SM00413E"}

@article{threecomp,
author = {Boyer, Franck and Lapuerta, Céline},
year = {2006},
month = {07},
pages = {653-687},
title = {Study of a three component Cahn-Hilliard flow model},
volume = {40},
journal = {Mathematical Modelling and Numerical Analysis},
doi = {10.1051/m2an:2006028}
}

@article{boyerncomp,
author = {Boyer, Franck and Minjeaud, Sebastian},
year = {2014},
month = {12},
pages = {2885-2928},
title = {Hierarchy of consistent n-component Cahn–Hilliard systems},
volume = {24},
journal = {Mathematical Models and Methods in Applied Sciences},
doi = {10.1142/S0218202514500407}
}

@article{dongncomp,
title = {Multiphase flows of N immiscible incompressible fluids: A reduction-consistent and thermodynamically-consistent formulation and associated algorithm},
journal = {Journal of Computational Physics},
volume = {361},
pages = {1-49},
year = {2018},
issn = {0021-9991},
doi = {https://doi.org/10.1016/j.jcp.2018.01.041},
author = {S. Dong}
}

@article{dongwall,
title = {Wall-bounded multiphase flows of N immiscible incompressible fluids: Consistency and contact-angle boundary condition},
journal = {Journal of Computational Physics},
volume = {338},
pages = {21-67},
year = {2017},
issn = {0021-9991},
doi = {https://doi.org/10.1016/j.jcp.2017.02.048},
url = {https://www.sciencedirect.com/science/article/pii/S0021999117301511},
author = {S. Dong}
}

@book{lbmbook1,
title = "The Lattice Boltzmann Method: Principles and Practice",
author = "Timm Krueger and Halim Kusumaatmaja and Alexandr Kuzmin and Orest Shardt and Goncalo Silva and Viggen, {Erlend Magnus}",
year = "2016",
series = "Graduate Texts in Physics",
publisher = "Springer",
}

@book{lbmbook2,
    author = {Succi, Sauro},
    title = {The Lattice Boltzmann Equation for Fluid Dynamics and Beyond},
    publisher = {Oxford University Press},
    year = {2001},
    month = {06},
    isbn = {9780198503989},
    doi = {10.1093/oso/9780198503989.001.0001},
    url = {https://doi.org/10.1093/oso/9780198503989.001.0001},
}

@article{lbmncomp2,
    author = {Yuan, Xiaolei and Shi, Baochang and Zhan, Chengjie and Chai, Zhenhua},
    title = {A phase-field-based lattice Boltzmann model for multiphase flows involving N immiscible incompressible fluids},
    journal = {Physics of Fluids},
    volume = {34},
    number = {2},
    pages = {023311},
    year = {2022},
    month = {02},
    issn = {1070-6631},
    doi = {10.1063/5.0078507}
}

@article{lbmncomp1,
title = {Reduction-consistent Cahn–Hilliard theory based lattice Boltzmann equation method for N immiscible incompressible fluids},
journal = {Physica A: Statistical Mechanics and its Applications},
volume = {574},
pages = {126015},
year = {2021},
issn = {0378-4371},
doi = {https://doi.org/10.1016/j.physa.2021.126015},
author = {Lin Zheng and Song Zheng and Qinglan Zhai}
}

@article{PhysRevE.97.033312,
  title = {Numerical simulation of three-component multiphase flows at high density and viscosity ratios using lattice Boltzmann methods},
  author = {Haghani Hassan Abadi, Reza and Fakhari, Abbas and Rahimian, Mohammad Hassan},
  journal = {Physical Review E},
  volume = {97},
  issue = {3},
  pages = {033312},
  numpages = {15},
  year = {2018},
  month = {Mar},
  publisher = {American Physical Society},
  doi = {10.1103/PhysRevE.97.033312}
}

@article{wellbalanced2,
    author = {Zhang, Chunhua and Guo, Zhaoli and Wang, Lian-Ping},
    title = {Improved well-balanced free-energy lattice Boltzmann model for two-phase flow with high Reynolds number and large viscosity ratio},
    journal = {Physics of Fluids},
    volume = {34},
    number = {1},
    pages = {012110},
    year = {2022},
    month = {01},
    issn = {1070-6631},
    doi = {10.1063/5.0072221},
}

@article{wellbalanced1,
  title = {Implementation of contact line motion based on the phase-field lattice Boltzmann method},
  author = {Ju, Long and Guo, Zhaoli and Yan, Bicheng and Sun, Shuyu},
  journal = {Physical Review E},
  volume = {109},
  issue = {4},
  pages = {045307},
  numpages = {15},
  year = {2024},
  month = {Apr},
  publisher = {American Physical Society},
  doi = {10.1103/PhysRevE.109.045307}
}

@article{
zwicker,
author = {David Zwicker  and Liedewij Laan },
title = {Evolved interactions stabilize many coexisting phases in multicomponent liquids},
journal = {Proceedings of the National Academy of Sciences},
volume = {119},
number = {28},
pages = {e2201250119},
year = {2022},
doi = {10.1073/pnas.2201250119},
URL = {https://www.pnas.org/doi/abs/10.1073/pnas.2201250119}}

@article{
chaderjian2025diversedistinctdenselypacked,
author = {Aria S. Chaderjian  and Sam Wilken  and Omar A. Saleh },
title = {Diverse, distinct, and densely packed DNA nanostar droplets},
journal = {Proceedings of the National Academy of Sciences},
volume = {123},
number = {7},
pages = {e2523462123},
year = {2026},
doi = {10.1073/pnas.2523462123},
URL = {https://www.pnas.org/doi/abs/10.1073/pnas.2523462123}}

@article{Wagner_2001,
doi = {10.1209/epl/i2001-00551-4},
url = {https://doi.org/10.1209/epl/i2001-00551-4},
year = {2001},
month = {nov},
publisher = {},
volume = {56},
number = {4},
pages = {556},
author = {A. J. Wagner and M. E. Cates},
title = {Phase ordering of
two-dimensional symmetric 
 binary fluids: A droplet scaling state},
journal = {Europhysics Letters}
}

@article{Furukawa,
  title = {Spinodal decomposition of two-dimensional fluid mixtures: A spectral analysis of droplet growth},
  author = {Furukawa, H.},
  journal = {Physical Review E},
  volume = {61},
  issue = {2},
  pages = {1423--1431},
  numpages = {0},
  year = {2000},
  month = {Feb},
  publisher = {American Physical Society},
  doi = {10.1103/PhysRevE.61.1423},
  url = {https://link.aps.org/doi/10.1103/PhysRevE.61.1423}
}

@article{WagnerYeomans,
  title = {Breakdown of Scale Invariance in the Coarsening of Phase-Separating Binary Fluids},
  author = {Wagner, Alexander J. and Yeomans, J. M.},
  journal = {Physical Review Letters},
  volume = {80},
  issue = {7},
  pages = {1429--1432},
  numpages = {0},
  year = {1998},
  month = {Feb},
  publisher = {American Physical Society},
  doi = {10.1103/PhysRevLett.80.1429},
  url = {https://link.aps.org/doi/10.1103/PhysRevLett.80.1429}
}

@Article{blockcopolymer,
author ="Mai, Yiyong and Eisenberg, Adi",
title  ="Self-assembly of block copolymers",
journal  ="Chemical Society Reviews",
year  ="2012",
volume  ="41",
issue  ="18",
pages  ="5969-5985",
publisher  ="The Royal Society of Chemistry",
doi  ="10.1039/C2CS35115C",
url  ="http://dx.doi.org/10.1039/C2CS35115C"}

@article{Zheng2015,
  title = {Lattice Boltzmann equation method for the Cahn-Hilliard equation},
  author = {Zheng, Lin and Zheng, Song and Zhai, Qinglan},
  journal = {Physical Review E},
  volume = {91},
  issue = {1},
  pages = {013309},
  numpages = {7},
  year = {2015},
  month = {Jan},
  publisher = {American Physical Society},
  doi = {10.1103/PhysRevE.91.013309},
  url = {https://link.aps.org/doi/10.1103/PhysRevE.91.013309}
}

@article{
xitong,
author = {Xitong Zhang  and Hongyu Zhao  and Jack R. Panter  and Glen McHale  and Gary G. Wells  and Rodrigo Ledesma-Aguilar  and Halim Kusumaatmaja },
title = {Multiple equilibria enables tunable wetting of droplets on patterned liquid surfaces},
journal = {Science Advances},
volume = {11},
number = {38},
pages = {eadw6615},
year = {2025},
doi = {10.1126/sciadv.adw6615},
URL = {https://www.science.org/doi/abs/10.1126/sciadv.adw6615}}

@article{Lee2010,
title = {Lattice Boltzmann simulations of micron-scale drop impact on dry surfaces},
journal = {Journal of Computational Physics},
volume = {229},
number = {20},
pages = {8045-8063},
year = {2010},
issn = {0021-9991},
doi = {https://doi.org/10.1016/j.jcp.2010.07.007},
url = {https://www.sciencedirect.com/science/article/pii/S0021999110003761},
author = {Taehun Lee and Lin Liu}
}

@book{rowlinson1982molecular,
  title        = {Molecular Theory of Capillarity},
  author       = {Rowlinson, J. S. and Widom, B.},
  year         = {1982},
  publisher    = {Clarendon Press},
  address      = {Oxford, UK},
  isbn         = {9780198556422}
}

@article{Postma,
  title = {Force methods for the two-relaxation-times lattice Boltzmann},
  author = {Postma, Bart and Silva, Goncalo},
  journal = {Physical Review E},
  volume = {102},
  issue = {6},
  pages = {063307},
  numpages = {7},
  year = {2020},
  month = {Dec},
  publisher = {American Physical Society},
  doi = {10.1103/PhysRevE.102.063307},
  url = {https://link.aps.org/doi/10.1103/PhysRevE.102.063307}
}

@article{Convective,
  title = {Evaluation of outflow boundary conditions for two-phase lattice Boltzmann equation},
  author = {Lou, Qin and Guo, Zhaoli and Shi, Baochang},
  journal = {Physical Review E},
  volume = {87},
  issue = {6},
  pages = {063301},
  numpages = {16},
  year = {2013},
  month = {Jun},
  publisher = {American Physical Society},
  doi = {10.1103/PhysRevE.87.063301},
  url = {https://link.aps.org/doi/10.1103/PhysRevE.87.063301}
}

@article{janusmicrofluidic,
author = {Kamperman, Tom and Trikalitis, Vasileios D. and Karperien, Marcel and Visser, Claas Willem and Leijten, Jeroen},
title = {Ultrahigh-Throughput Production of Monodisperse and Multifunctional Janus Microparticles Using in-Air Microfluidics},
journal = {ACS Applied Materials \& Interfaces},
volume = {10},
number = {28},
pages = {23433-23438},
year = {2018},
doi = {10.1021/acsami.8b05227},

URL = { 
    
        https://doi.org/10.1021/acsami.8b05227
    
    

},

}

@Article{Naga2024,
author={Naga, Abhinav
and Rennick, Michael
and Hauer, Lukas
and Wong, William S. Y.
and Sharifi-Aghili, Azadeh
and Vollmer, Doris
and Kusumaatmaja, Halim},
title={Direct visualization of viscous dissipation and wetting ridge geometry on lubricant-infused surfaces},
journal={Communications Physics},
year={2024},
month={Sep},
day={17},
volume={7},
number={1},
pages={306},
issn={2399-3650},
doi={10.1038/s42005-024-01795-3},
url={https://doi.org/10.1038/s42005-024-01795-3}
}

@article{keiser_drop_2017,
	title = {Drop friction on liquid-infused materials},
	volume = {13},
	doi = {10.1039/C7SM01226H},
	number = {39},
	journal = {Soft Matter},
	author = {Keiser, Armelle and Keiser, Ludovic and Clanet, Christophe and Quéré, David},
	year = {2017},
	pages = {6981--6987},
}

@article{daniel_oleoplaning_2017,
	title = {Oleoplaning droplets on lubricated surfaces},
	volume = {13},
	doi = {10.1038/nphys4177},
	number = {10},
	journal = {Nature Physics},
	author = {Daniel, Dan and Timonen, Jaakko V. I. and Li, Ruoping and Velling, Seneca J. and Aizenberg, Joanna},
	year = {2017},
	pages = {1020--1025},
}

@article{microfluidic5, title={Modelling double emulsion formation in planar flow-focusing microchannels}, volume={895}, DOI={10.1017/jfm.2020.299}, journal={Journal of Fluid Mechanics}, author={Wang, Ningning and Semprebon, Ciro and Liu, Haihu and Zhang, Chuhua and Kusumaatmaja, Halim}, year={2020}, pages={A22}}

@article{microreactor1,
author = {Jia, Yankai and Ren, Yukun and Hou, Likai and Liu, Weiyu and Deng, Xiaokang and Jiang, Hongyuan},
title = {Sequential Coalescence Enabled Two-Step Microreactions in Triple-Core Double-Emulsion Droplets Triggered by an Electric Field},
journal = {Small},
volume = {13},
number = {46},
pages = {1702188},
doi = {https://doi.org/10.1002/smll.201702188},
url = {https://onlinelibrary.wiley.com/doi/abs/10.1002/smll.201702188},
year = {2017}
}

@article{microreactor2,
author = {Shum, Ho Cheung and Bandyopadhyay, Amit and Bose, Susmita and Weitz, David A.},
title = {Double Emulsion Droplets as Microreactors for Synthesis of Mesoporous Hydroxyapatite},
journal = {Chemistry of Materials},
volume = {21},
number = {22},
pages = {5548-5555},
year = {2009},
doi = {10.1021/cm9028935},

URL = { 
    
        https://doi.org/10.1021/cm9028935
    
    

},

}

@Article{lisangle1,
author ="Schellenberger, Frank and Xie, Jing and Encinas, Noemí and Hardy, Alexandre and Klapper, Markus and Papadopoulos, Periklis and Butt, Hans-Jürgen and Vollmer, Doris",
title  ="Direct observation of drops on slippery lubricant-infused surfaces",
journal  ="Soft Matter",
year  ="2015",
volume  ="11",
issue  ="38",
pages  ="7617-7626",
publisher  ="The Royal Society of Chemistry",
doi  ="10.1039/C5SM01809A",
url  ="http://dx.doi.org/10.1039/C5SM01809A"}

@article{lisangle2,
author = {Bottone, Davide and Seeger, Stefan},
title = {Droplet Memory on Liquid-Infused Surfaces},
journal = {Langmuir},
volume = {39},
number = {17},
pages = {6160-6168},
year = {2023},
doi = {10.1021/acs.langmuir.3c00289},


URL = { 
    
        https://doi.org/10.1021/acs.langmuir.3c00289
    
    

},


}

@article{diffuse1,
   author = "Anderson, D. M. and McFadden, G. B. and Wheeler, A. A.",
   title = "DIFFUSE-INTERFACE METHODS IN FLUID MECHANICS", 
   journal= "Annual Review of Fluid Mechanics",
   year = "1998",
   volume = "30",
   number = "Volume 30, 1998",
   pages = "139-165",
   doi = "https://doi.org/10.1146/annurev.fluid.30.1.139",
   url = "https://www.annualreviews.org/content/journals/10.1146/annurev.fluid.30.1.139",
   publisher = "Annual Reviews",
   issn = "1545-4479",
   type = "Journal Article",
  }

@article{diffuse2,
  title = {Ternary free-energy lattice Boltzmann model with tunable surface tensions and contact angles},
  author = {Semprebon, Ciro and Kr\"uger, Timm and Kusumaatmaja, Halim},
  journal = {Phys. Rev. E},
  volume = {93},
  issue = {3},
  pages = {033305},
  numpages = {11},
  year = {2016},
  month = {Mar},
  publisher = {American Physical Society},
  doi = {10.1103/PhysRevE.93.033305},
  url = {https://link.aps.org/doi/10.1103/PhysRevE.93.033305}
}

@article{SUSSMAN2007469,
title = {A sharp interface method for incompressible two-phase flows},
journal = {Journal of Computational Physics},
volume = {221},
number = {2},
pages = {469-505},
year = {2007},
issn = {0021-9991},
doi = {https://doi.org/10.1016/j.jcp.2006.06.020},
url = {https://www.sciencedirect.com/science/article/pii/S0021999106002981},
author = {M. Sussman and K.M. Smith and M.Y. Hussaini and M. Ohta and R. Zhi-Wei}
}

@article{BRACKBILL1992335,
title = {A continuum method for modeling surface tension},
journal = {Journal of Computational Physics},
volume = {100},
number = {2},
pages = {335-354},
year = {1992},
issn = {0021-9991},
doi = {https://doi.org/10.1016/0021-9991(92)90240-Y},
url = {https://www.sciencedirect.com/science/article/pii/002199919290240Y},
author = {J.U Brackbill and D.B Kothe and C Zemach}
}

@article{ABUALSAUD2018896,
title = {A conservative and well-balanced surface tension model},
journal = {Journal of Computational Physics},
volume = {371},
pages = {896-913},
year = {2018},
issn = {0021-9991},
doi = {https://doi.org/10.1016/j.jcp.2018.02.022},
url = {https://www.sciencedirect.com/science/article/pii/S0021999118301049},
author = {Moataz O. Abu-Al-Saud and Stéphane Popinet and Hamdi A. Tchelepi}
}

@article{microshear,
author = {Zhu, Zhiqiang and Huang, Fangsheng and Yang, Chaoyu and Si, Ting and Xu, Ronald X.},
title = {On-Demand Generation of Double Emulsions Based on Interface Shearing for Controlled Ultrasound Activation},
journal = {ACS Applied Materials \& Interfaces},
volume = {11},
number = {43},
pages = {40932-40943},
year = {2019},
doi = {10.1021/acsami.9b15182},


URL = { 
    
        https://doi.org/10.1021/acsami.9b15182
    
    

}

}

@Article{Lang2022,
author={Lang, Chao
and Lloyd, Elisabeth C.
and Matuszewski, Kelly E.
and Xu, Yifan
and Ganesan, Venkat
and Huang, Rui
and Kumar, Manish
and Hickey, Robert J.},
title={Nanostructured block copolymer muscles},
journal={Nature Nanotechnology},
year={2022},
month={Jul},
day={01},
volume={17},
number={7},
pages={752-758},
issn={1748-3395},
doi={10.1038/s41565-022-01133-0},
url={https://doi.org/10.1038/s41565-022-01133-0}
}

@Article{Peinemann2007,
author={Peinemann, Klaus-Viktor
and Abetz, Volker
and Simon, Peter F. W.},
title={Asymmetric superstructure formed in a block copolymer via phase separation},
journal={Nature Materials},
year={2007},
month={Dec},
day={01},
volume={6},
number={12},
pages={992-996},
issn={1476-4660},
doi={10.1038/nmat2038},
url={https://doi.org/10.1038/nmat2038}
}

@Article{cosmetics3040037,
AUTHOR = {Kulkarni, Chandrashekhar V.},
TITLE = {Lipid Self-Assemblies and Nanostructured Emulsions for Cosmetic Formulations},
JOURNAL = {Cosmetics},
VOLUME = {3},
YEAR = {2016},
NUMBER = {4},
ARTICLE-NUMBER = {37},
URL = {https://www.mdpi.com/2079-9284/3/4/37},
ISSN = {2079-9284}
}

@article{KLOJDOVA2022100451,
title = {W/o/w multiple emulsions: A novel trend in functional ice cream preparations?},
journal = {Food Chemistry: X},
volume = {16},
pages = {100451},
year = {2022},
issn = {2590-1575},
doi = {https://doi.org/10.1016/j.fochx.2022.100451},
url = {https://www.sciencedirect.com/science/article/pii/S2590157522002498},
author = {Iveta Klojdová and Constantinos Stathopoulos}
}

@Article{Lareau2019,
author={Lareau, Caleb A.
and Duarte, Fabiana M.
and Chew, Jennifer G.
and Kartha, Vinay K.
and Burkett, Zach D.
and Kohlway, Andrew S.
and Pokholok, Dmitry
and Aryee, Martin J.
and Steemers, Frank J.
and Lebofsky, Ronald
and Buenrostro, Jason D.},
title={Droplet-based combinatorial indexing for massive-scale single-cell chromatin accessibility},
journal={Nature Biotechnology},
year={2019},
month={Aug},
day={01},
volume={37},
number={8},
pages={916-924},
issn={1546-1696},
url={https://doi.org/10.1038/s41587-019-0147-6}
}

@Article{Zheng2017,
author={Zheng, Grace X. Y.
and Terry, Jessica M.
and Belgrader, Phillip
and Ryvkin, Paul
and Bent, Zachary W.
and Wilson, Ryan
and Ziraldo, Solongo B.
and Wheeler, Tobias D.
and McDermott, Geoff P.
and Zhu, Junjie
and Gregory, Mark T.
and Shuga, Joe
and Montesclaros, Luz
and Underwood, Jason G.
and Masquelier, Donald A.
and Nishimura, Stefanie Y.
and Schnall-Levin, Michael
and Wyatt, Paul W.
and Hindson, Christopher M.
and Bharadwaj, Rajiv
and Wong, Alexander
and Ness, Kevin D.
and Beppu, Lan W.
and Deeg, H. Joachim
and McFarland, Christopher
and Loeb, Keith R.
and Valente, William J.
and Ericson, Nolan G.
and Stevens, Emily A.
and Radich, Jerald P.
and Mikkelsen, Tarjei S.
and Hindson, Benjamin J.
and Bielas, Jason H.},
title={Massively parallel digital transcriptional profiling of single cells},
journal={Nature Communications},
year={2017},
month={Jan},
day={16},
volume={8},
number={1},
pages={14049},
issn={2041-1723},
url={https://doi.org/10.1038/ncomms14049}
}

@article{Raeli2025,
    author = {Raeli, Alice and Borello, Eloisa Salina and Serazio, Cristina and Czelusniak, Luiz Eduardo and Bingert, Tim Niklas and Krause, Mathias J. and Viberti, Dario},

    title = {Analysis of Lattice Boltzmann Method Potentials for Understanding Underground Fluid Storage Microscale Phenomena},

    volume = {OMC Med Energy Conference and Exhibition},

    series = {Offshore Mediterranean Conference and Exhibition},

    pages = {OMC-2025-597},

    year = {2025},

    month = {04},


}

@software{kummerlander_2025_17899765,
  author       = {Kummerländer, Adrian and
                  Bingert, Tim and
                  Bock, Simon and
                  Bukreev, Fedor and
                  Castroviejo, Daniel and
                  Czelusniak, Luiz Eduardo and
                  Dapelo, Davide and
                  Gaul, Christoph and
                  Dorn, Marcio and
                  Dorneles, Leonardo and
                  Grafen, Johannes and
                  Grinschewski, Michael and
                  Ito, Shota and
                  Jeßberger, Julius and
                  Kaiser, Florian and
                  Khazaeipoul, Danial and
                  Krüger, Timm and
                  Kumbhat, Arsh and
                  Kusumaatmaja, Halim and
                  Nettekoven, Andreas and
                  Raeli, Alice and
                  Riazantsev, Tikhon and
                  Rennick, Michael and
                  Prakash, Gagan and
                  Prinz, František and
                  Sauterleute, Liam and
                  Schecher, Maximilian and
                  Schneider, Andreas and
                  Shimojima, Yuji and
                  Simonis, Stephan and
                  Spelten, Philipp and
                  Tacques, Alexandre and
                  Krause, Mathias J.},
  title        = {OpenLB Release 1.9: Open Source Lattice Boltzmann
                   Code
                  },
  month        = dec,
  year         = 2025,
  publisher    = {Zenodo},
  version      = {1.9.0},
  doi          = {10.5281/zenodo.17899765},
  url          = {https://doi.org/10.5281/zenodo.17899765},
}

@Article{Fabrini2024,
author={Fabrini, Giacomo
and Farag, Nada
and Nuccio, Sabrina Pia
and Li, Shiyi
and Stewart, Jaimie Marie
and Tang, Anli A.
and McCoy, Reece
and Owens, R{\'o}is{\'i}n M.
and Rothemund, Paul W. K.
and Franco, Elisa
and Di Antonio, Marco
and Di Michele, Lorenzo},
title={Co-transcriptional production of programmable RNA condensates and synthetic organelles},
journal={Nature Nanotechnology},
year={2024},
month={Nov},
day={01},
volume={19},
number={11},
pages={1665-1673},
issn={1748-3395},
doi={10.1038/s41565-024-01726-x},
url={https://doi.org/10.1038/s41565-024-01726-x}
}

@ARTICLE{Abraham2024-io,
  title    = "Nucleic acid liquids",
  author   = "Abraham, Gabrielle R and Chaderjian, Aria S and N Nguyen, Anna B
              and Wilken, Sam and Saleh, Omar A",
  journal  = "Rep Prog Phys",
  volume   =  87,
  number   =  6,
  month    =  may,
  year     =  2024,
  address  = "England",
  language = "en"
}

@article{olbPaper2021,

  author   = {Krause, M.J. and Kummerl\"ander, A. and Avis, S.J. and Kusumaatmaja, H. and Dapelo, D. and Klemens, F. and Gaedtke, M. and Hafen, N. and Mink, A. and Trunk, R. and Marquardt, J.E. and Maier, M.L. and Haussmann, M. and Simonis, S.},

  title    = {{OpenLB--Open source lattice Boltzmann code}},

  doi      = {https://doi.org/10.1016/j.camwa.2020.04.033},

  issn     = {0898-1221},

  pages    = {258--288},

  url      = {http://www.sciencedirect.com/science/article/pii/S0898122120301875},

  volume   = {81},

  journal  = {Computers \& Mathematics with Applications},

  year     = {2021},

}

@article{Ladd_1994, title={Numerical simulations of particulate suspensions via a discretized Boltzmann equation. Part 1. Theoretical foundation}, volume={271}, DOI={10.1017/S0022112094001771}, journal={Journal of Fluid Mechanics}, author={Ladd, Anthony J. C.}, year={1994}, pages={285–309}}

@article{TIRIBOCCHI20251,
title = {Lattice Boltzmann simulations for soft flowing matter},
journal = {Physics Reports},
volume = {1105},
pages = {1-52},
year = {2025},
note = {Lattice Boltzmann simulations for soft flowing matter},
issn = {0370-1573},
doi = {https://doi.org/10.1016/j.physrep.2024.11.002},
url = {https://www.sciencedirect.com/science/article/pii/S0370157324003831},
author = {Adriano Tiribocchi and Mihir Durve and Marco Lauricella and Andrea Montessori and Jean-Michel Tucny and Sauro Succi}
}

@article{Montessori_Lauricella_Tirelli_Succi_2019, title={Mesoscale modelling of near-contact interactions for complex flowing interfaces}, volume={872}, DOI={10.1017/jfm.2019.372}, journal={Journal of Fluid Mechanics}, author={Montessori, A. and Lauricella, M. and Tirelli, N. and Succi, S.}, year={2019}, pages={327–347}}

@article{ALamura_1999,
doi = {10.1209/epl/i1999-00165-4},
url = {https://doi.org/10.1209/epl/i1999-00165-4},
year = {1999},
month = {feb},
publisher = {},
volume = {45},
number = {3},
pages = {314},
author = {A. Lamura and G. Gonnella and J. M. Yeomans},
title = {A lattice Boltzmann model of ternary fluid mixtures},
journal = {Europhysics Letters}
}

@article{abadi_ternary_2018,
  title = {Numerical simulation of three-component multiphase flows at high density and viscosity ratios using lattice Boltzmann methods},
  author = {Haghani Hassan Abadi, Reza and Fakhari, Abbas and Rahimian, Mohammad Hassan},
  journal = {Physical Review E},
  volume = {97},
  issue = {3},
  pages = {033312},
  numpages = {15},
  year = {2018},
  month = {Mar},
  publisher = {American Physical Society},
  doi = {10.1103/PhysRevE.97.033312},
  url = {https://link.aps.org/doi/10.1103/PhysRevE.97.033312}
}

\end{document}